\documentclass[11pt, letterpaper]{article}
\usepackage[utf8]{inputenc}
\usepackage{geometry}
\usepackage{amsmath}
\usepackage{amssymb}
\usepackage{graphicx}
\usepackage{bm}
\usepackage{mathtools}
\usepackage{hyperref}
\usepackage{tikz}
\usepackage{setspace}
\usepackage[round]{natbib}
\usepackage{subcaption}
\usepackage{bbm}
\usepackage{mathbbol}
\usepackage{xcolor}
\usepackage{pdfpages}

\def\spacingset#1{\renewcommand{\baselinestretch}%
{#1}\small\normalsize} \spacingset{1}

\newcommand{\blind}{0}

\usepackage{stackengine}
\usepackage{rotating}
\usepackage{booktabs}
\usepackage{array}
\usepackage{multirow}
\usepackage{wrapfig}
\usepackage{float}
\usepackage{colortbl}
\usepackage{pdflscape}
\usepackage{tabu}
\usepackage{threeparttable}
\usepackage{threeparttablex}
\usepackage[normalem]{ulem}
\usepackage{makecell}
\usepackage{xcolor}

\usetikzlibrary{arrows, automata}
\usepackage{fullpage}

\renewcommand{\arraystretch}{1.3}

\usepackage{lmodern}
\usepackage{amssymb,amsmath}
\usepackage{ifxetex,ifluatex}
\ifnum 0\ifxetex 1\fi\ifluatex 1\fi=0 
  \usepackage[T1]{fontenc}
  \usepackage[utf8]{inputenc}
  \usepackage{textcomp} 
\else 
  \usepackage{unicode-math}
  \defaultfontfeatures{Scale=MatchLowercase}
  \defaultfontfeatures[\rmfamily]{Ligatures=TeX,Scale=1}
\fi
\IfFileExists{upquote.sty}{\usepackage{upquote}}{}
\IfFileExists{microtype.sty}{
  \usepackage[]{microtype}
  \UseMicrotypeSet[protrusion]{basicmath} 
}{}
\makeatletter
\@ifundefined{KOMAClassName}{
  \IfFileExists{parskip.sty}{%
    \usepackage{parskip}
  }{
    \setlength{\parindent}{0pt}
    \setlength{\parskip}{6pt plus 2pt minus 1pt}}
}{
  \KOMAoptions{parskip=half}}
\makeatother
\usepackage{xcolor}
\IfFileExists{xurl.sty}{\usepackage{xurl}}{} 
\IfFileExists{bookmark.sty}{\usepackage{bookmark}}{\usepackage{hyperref}}
\hypersetup{
  pdftitle={KPop Simulations on stats cluster},
  hidelinks,
  pdfcreator={LaTeX via pandoc}}
\usepackage{fancyvrb}

\DefineVerbatimEnvironment{Highlighting}{Verbatim}{commandchars=\\\{\}}
\usepackage{framed}

\definecolor{shadecolor}{RGB}{248,248,248}

\usepackage{graphicx,grffile}
\makeatletter
\def\maxwidth{\ifdim\Gin@nat@width>\linewidth\linewidth\else\Gin@nat@width\fi}
\def\maxheight{\ifdim\Gin@nat@height>\textheight\textheight\else\Gin@nat@height\fi}
\makeatother
\setkeys{Gin}{width=\maxwidth,height=\maxheight,keepaspectratio}
\makeatletter
\def\fps@figure{htbp}
\makeatother
\usepackage{standalone}

\newcommand{\indep}{\rotatebox[origin=c]{90}{$\models$}}
\newcommand{\E}{\mathbbm{E}}
\newcommand{\X}{\mathbf{X}}
\newcommand{\K}{\mathbf{K}}
\newcommand{\Y}{\mathbf{Y}}
\newcommand{\U}{\mathbf{U}}
\newcommand{\V}{\mathbf{V}}
\newcommand{\A}{\mathbf{A}}

\newcommand{\Ns}{N_{s}}
\newcommand{\Np}{N_{pop}}
\newcommand{\kpop}{\text{\emph{kpop}}}

\DeclareMathOperator*{\argmin}{arg\,min}

\newtheorem{estimator}{\small\sc Estimator}
\newtheorem{proposition}{\small\sc Proposition}
\newtheorem{assumption}{\small\sc Assumption}

\begin{document}
\setcounter{page}{0}
\thispagestyle{empty}

\if0\blind
{
  \title{\bf \kpop: A kernel balancing approach for reducing specification assumptions in survey weighting\thanks{
    The authors thank Avi Feller, Kosuke Imai, Luke Miratrix, Santiago Olivella, Alex Tarr, Baobao Zhang and participants in the UCLA Causal Inference Reading Group.  This work is partially supported by a grant from Facebook Statistics for Improving Insights and Decisions. The \texttt{kbal} package for the \texttt{R} computing language implements this method and is freely available.}}
  \footnotesize
 \author{Erin Hartman\thanks{Assistant Professor, Department of Political Science, University of California, Berkeley (corresponding). Email: \href{mailto:ekhartman@berkeley.edu}{ekhartman@berkeley.edu} URL:
     \href{http://www.erinhartman.com}{http://www.erinhartman.com}.}  \hspace{1in}
    Chad Hazlett\thanks{Professor, Department of Statistics \& Data Science, Department of Political Science, University of California Los Angeles (corresponding). Email: \href{mailto:chazlett@ucla.edu}{chazlett@ucla.edu} URL:
     \href{http://www.chadhazlett.com}{http://www.chadhazlett.com}.}
     \hspace{1in}
     Ciara Sterbenz\thanks{PhD Candidate, Department of Political Science, University of California Los Angeles. \newline
     Email:
     \href{mailto:cster@ucla.edu}{cster@ucla.edu}.}}
  \maketitle
} \fi

\if1\blind
{
  \bigskip
  \bigskip
  \bigskip
  \begin{center}
    {\LARGE\bf \kpop: A kernel balancing approach for reducing specification assumptions in survey weighting}
\end{center}
  \medskip
} \fi

\bigskip
\begin{abstract} 
With the precipitous decline in response rates, researchers and pollsters have been left with highly non-representative samples, relying on constructed weights to make these samples representative of the desired target population. Though practitioners employ valuable expert knowledge to choose what variables, $X$ must be adjusted for, they rarely defend particular functional forms relating these variables to the response process or the outcome. Unfortunately, commonly-used calibration weights---which make the weighted mean $X$ in the sample equal that of the population---only ensure correct adjustment when the portion of the outcome and the response process left unexplained by linear functions of $X$ are independent. To alleviate this functional form dependency, we describe kernel balancing for population weighting (\kpop). This approach replaces the design matrix $\X$ with a kernel matrix, $\K$ encoding high-order information about $\X$. Weights are then found to make the weighted average row of $\K$ among sampled units approximately equal that of the target population. This produces good calibration on a wide range of smooth functions of $X$, without relying on the user to decide which $X$ or what functions of them to include. We describe the method and illustrate it by application to polling data from the 2016 U.S. presidential election.
\end{abstract}

\noindent%
\vfill

\pagebreak
\clearpage

\spacingset{1.5} 

\newpage
\setcounter{page}{1}

\clearpage
 
\newpage
\setcounter{page}{1}

\section{Introduction}

In an era of decreasing response rates, social scientists must rely on methods to adjust for the non-representative nature of survey samples.  For example, Pew Research Center saw response rates to live-caller phone surveys decline from nearly one third of respondents in the late 1990s, to only 6\% in 2018 \citep{kennedy_hartig_2019}.  The non-random nature of this ``unit nonresponse'' poses serious challenges for survey researchers and has led to greater use of non-probability sampling methods, such as panel, quota, or river sampling for online surveys \citep{Mercer:2017gb}.  The concern, whether due to nonresponse or non-probability sampling, is that the resulting survey respondents are not representative of the target population about which a researcher aims to draw an inference, leaving the potential for significant bias in estimates of target outcomes.

Researchers are therefore often obligated to construct survey weights to address this bias.
Constructing these weights requires researchers to choose (1) what variables to account for in the weighting procedure, and (2) how to incorporate these variables in the construction of survey weights? For example, researchers have determined  pollsters' failure to account for education level in survey weighting resulted in inaccurate predictions leading up to the 2016 U.S. Presidential election.  Even those that did account for educational attainment often failed to account for low levels of Midwestern, white voters with lower levels of educational attainment, i.e. the interaction of region, race, and education level \citep{kennedy2018evaluation}.  We return to this issue in our application, demonstrating how our proposed method can address these concerns. 

We begin with the observation that in practice, researchers seek to make the sample and target population identical only on some summary of the characteristics, represented by the matrix $\X$. The variables taking the columns in $\X$ may include indicators for membership in intersectional strata, and/or the values of other variables. In practice, $\X$ is typically (i) low-dimensional and (ii) chosen or constructed by hand. Weights are then chosen to make the sample similar to the target population in terms of the means of these $\X$,  thereby neglecting other moments of $p(\X)$ that, unnoticed, can remain dissimilar between the weighted sample and target population. 

Unfortunately, except in the case of full saturation (i.e. every combination of $\X$ values can be represented by an indicator), investigators are not generally in a position to argue that the outcome or response process are linear in such $\X$, which is needed to achieve unbiasedness through this adjustment (see e.g. \citealp{sarndal2005estimation, Kott:2010dq}, with analogous results in the causal inference setting, see e.g. \citealp{zhao2016entropy}). Note that  ``response'', in this context, means that a participant was both sampled and responded and thus appears in the observed data. Further, leaving the choice of which variables and which higher-order terms to include in the hands of investigators allows almost unlimited researcher degrees of freedom. As we show, even across a set of seemingly reasonable choices, the resulting estimates can vary widely. 

The question is then how researchers can choose what functions of covariates, $\phi(\X)$, should be used for constructing weights. We provide one reasoned answer to this question, aiming to require the weakest workable assumptions and minimal user intervention. To do so first requires a clarification of how the non-parametric identification assumptions invoked to handle non-response become parametric assumptions once we are also constrained by estimation concerns. Specifically, we formulate the ``linear ignorability'' assumption, which states that survey weights unbiasedly estimate the desired outcome among the target population only when the part of the outcome not explained by a linear combination of the $\phi(\X)$ is independent of the part of the sampling process not explained linearly by $\phi(\X)$ within a suitable link function. As we detail below, this refines and weakens existing results that call for both non-parametric ignorability of selection and linearity of the outcome (or selection model) in $\phi(\X)$ as separate matters.  

Our main contribution is to propose a specific kernel-based weighting procedure (\kpop) as a practical estimation procedure that reduces bias by more nearly meeting the linear-ignorability assumption. In short, this approach employs the kernel matrix, $\K$, whose linear span captures a wide range of smooth, non-linear functions of $\X$. Weights are then chosen to make the weighted average row of $\K$ in the sample approximately equal to the average row of $\K$ in the target population. Weights chosen in this way approximately equate the (weighted) distribution of $\X$ in the survey with that of the target group, as would be estimated by a kernel density estimator \citep{hazlett2020kernel}.

In what follows, Section~\ref{sec:setting} establishes our setting and notation. Section~\ref{sec:estimationidentification} reviews calibration estimators and discusses identification, introducing a new minimal identification requirement. Section~\ref{sec:proposal} describes our proposed kernel based calibration estimator, while Section~\ref{sec:simulations} demonstrates its behavior in two simulated examples. 
We apply the technique to to the 2016 U.S. Presidential election in Section~\ref{sec:application}, showing how \kpop\ can be used to mitigate concerns due to limited foreknowledge of what interactions or intersectional strata are important. Section~\ref{sec:discussion} concludes.

\section{Setting and notation}\label{sec:setting}

Our setting considers two primary objects. The first is a sample of the form $\{X_i, Z_i, Y_i, R_i \}_{i=1}^{\Ns}$, where $Y_i$ is the outcome of interest, and $X_i$ is a collection of $P$ auxiliary variables (covariates) that will be adjusted for. The auxiliary data initially encoded as $X$ may be mapped to a richer feature expansion $\phi(X)$, with $X \mapsto \phi(X)$ from $\mathbb{R}^P \mapsto \mathbb{R}^{P'}$, potentially with $P'>>P$. In the survey setting, typically many or all dimensions of $X$ are categorical, as in education levels, party identification, etc. We will consider such cases here, though the methods described are equally natural for continuous variables. $R_i$ is an indicator for selection into the sample with  $R_i=1$ for all units in the sample. $Z_i$ is included in each tuple here to represent a potentially important unobserved factor, which will become important when considering the conditions that will lead to biased estimates. Each tuple in this sample is presumed to be drawn independently from an unknown joint density. 

The second object of interest is a larger but still finite target group or, in some contexts, finite population. This is a collection $\{X_i, Z_i, Y_i, R_i \}_{i=1}^{\Np}$, with $\Np>>\Ns$. Critically, each tuple in this collection is drawn from a common joint density $p(X,Z,Y,R)$. Depending on field and custom, this common joint density is sometimes referred to as the data generating process, the true population of interest, or the super-population from which the target group or population was drawn but from which many others could have in theory been drawn. Note that $X_i$ must be observed for all units in both groups to allow adjustment. However, $Y$ is unobserved in this target group. The immediate target of inference is the mean of $Y$ in this larger group. In some cases, this target group is identical with the ultimate target population of interest (e.g. when it is a census). In other cases, the mean of $Y$ over the larger group is of interest principally as an estimate of the expectation of $Y$ over $p()$. For example, the target may be a very large representative survey from a national population, as in our application. While this poses no problem in terms of bias (provided the survey is indeed representative of the referenced population), note that we do not consider here how one's uncertainty estimate would change when targeting the population mean rather than the mean over the observed target group (though see \citealp{opsomer2021replication}). In addition, context will determine whether the smaller survey sample is a subset of the larger, or if the smaller survey sample is independently or disjointly drawn. We proceed in the latter setting for congruence with our applied example and for notational simplicity. However, the former is easily accommodated and requires only small changes; see Appendix~\ref{app:proofs}.

The key problem to address is that, due to the nature of data collection---e.g. the possibility of selective response to the survey---the finite sample is drawn from 
a different distribution than the $p(X,Z,Y,R)$ describing the target group, and the user has at best incomplete knowledge as to how these distributions differ. Our goal is to estimate weights such that the weighted mean of $Y$ in the sample is the best possible estimate for the mean of $Y$ in the target group.

\section{Estimation and identification}\label{sec:estimationidentification}

\subsection{Calibration estimators}

Suppose researchers have only a few variables in $X_i$, and each is discrete with a small number of categories. In such settings it is straightforward to adjust for $X$ without any functional form or specification commitments: one can take the sample data, average the $Y$ within each stratum of $X$, then re-average these strata-wise averages together according to how often each stratum appeared in the target group. This is the \textit{post-stratification} estimator, and it can provide an unbiased estimate of the mean of $Y$ in the target group under the non-parametric identification assumption that conditional on $X$, $Y$ is independent of $R$ (conditional ignorability; see below). 

Unfortunately, such an approach is often infeasible because $X$ contains one or more continuous variables, and/or some strata that may be non-empty in the target group are empty in the sample. In such cases, investigators most often turn to calibration estimators, which construct weights on the sampled units such that the weighted mean of $\phi(X)$ among the sample equals the mean of  $\phi(X)$ among the target population, where $\phi(X)$ represents some chosen transformation of the original $X$. In general form, calibration weights $w$ are estimated according to:
\begin{align}\label{calibration}
    \min_{w} \; D(w, q) \quad \text{s.t.} \sum_{i : R_i = 1} w_i \phi(X_i) = T, \; \sum_{i : R_i = 1} w_i = 1, \text{\ and \ } 0 \leq w_i \leq 1
\end{align}

where $q_i$ refers to a reference or base weight and $D(\cdot, \cdot)$ corresponds to a distance or divergence metric, acting as a measure of how extremely the weights diverge from $q_i$. In principle $q_i$ may be the design weights for the sampling strategy employed. However, these are often unknown or unavailable, or far less influential after conditioning on the auxiliary variables. One may then let $q_i = 1/\Ns$ be the uniform base weight for units in the respondent sample. The vector $T$ describes target population moment constraints based on the mapping $\phi(X)$. Typically, and in our case, this is an average of $\phi(X)$ in the target population, which we treat as known, but which may be estimated, in which case that additional uncertainty should be propagated \citep{opsomer2021replication}. In other words,  the constraint $\sum_{i : R_i = 1} w_i \phi(X_i) = T$ in~\eqref{calibration} is the ``balance condition'' to be satisfied, 
\begin{equation}\label{eq:balance}
\sum_{i:R=1} w_{i} \phi(X_i) = \frac{1}{\Np}\sum_{j=1}^{\Np} \phi(X_j)  
\end{equation}

\noindent We note that meeting the conditions above,  particularly the balance condition, is not always feasible, and becomes less feasible as the dimensionality of $\phi(\cdot)$ grows. Addressing this will require some combination of relaxing the balance constraints (i.e. ``approximate balance''), or reducing the richness of $\phi(\cdot)$. These tradeoffs must be managed by any practical proposal. 

Common types of survey weighting correspond to different distance metrics $D(\cdot, \cdot)$, and are closely related to generalized regression estimation \citep{Sarndal:2007vj}.  We use $D(w, q) = \sum_{i: R = 1} w_i log (w_i / q_i)$, commonly employed in ``raking'' methods and variably known as exponential tilting \citep{Wu:2016ba}, maximum-entropy weighting, or entropy balancing \citep{hainmueller2012entropy}. Other distance metrics correspond to other common weighting estimators, although the choice of distance metric matters far less than the choice of moment constraints \cite{Deville:1992bv}.
For  broader reviews of calibration, see \citet{Sarndal:2007vj}, \citet{caughey_elements_2020}, or \citet{Wu:2016ba}. The constraints in~(\ref{calibration}) ensure the weights are non-negative and sum to one, ensuring they have a probability-like interpretation. Relaxing this constraint, e.g. allowing negative weights, would be to allow for ``extrapolation'' beyond the support of the respondent sample. This increases the possibility of severe model dependency, but is employed in some techniques such as generalized regression estimators  \citep{Deville:1992bv}.

Finally, the average outcome among the target population, $\mu = \frac{1}{\Np}\sum_i Y_i$, is then estimated as 
\begin{estimator}[Calibration Estimator for Target Population Mean of $Y$]\label{eqn:cal_estimator}
  \[\hat{\mu} = \sum_{i:R_i = 1} w_i Y_i\]
  with weights chosen by Equation~\eqref{calibration}
\end{estimator}

In principle, calibration is a general and powerful tool given the flexibility of the choice of $\phi(X)$. In practice, however, most applications of calibration simply seek to match the means of $X$ in the sample to that of the population, i.e. $\phi(X)=X$. We refer to this as \textit{mean calibration}, understanding that ``mean'' refers to the original $X$. Such an approach holds intuitive appeal since, at minimum, pollsters and investigators seek to adjust a sample to closely match a target population on the margins, particularly on variables such as the proportion falling in some demographic or partisan group. The risk, however, is that there is little reason to expect key identification assumption to hold under mean calibration---a problem we turn to now.

\subsection{Identification: From non-parametric ignorability to ``linear ignorability''}

Under what assumptions regarding the data generating process is it possible for the calibration estimator to unbiasedly estimate the average $Y$ in the target group? Typically this question is answered by appealing to the ignorability of the response conditionally on the observed covariates used for adjustment \citep{little2019statistical},
\begin{assumption}[Non-parametric ignorability of response]\label{assumption.npignorable}
$$Y \indep R \mid X $$
\end{assumption}

\noindent
The key shortcoming of this identification strategy is simply that investigators cannot typically invoke non-parametric conditioning in practice, and alternatives such as calibration end up calling for different or additional assumptions.  That is, supposing the target group and sample are disjoint, under non-parametric ignorability (Assumption~\ref{assumption.npignorable}) the target is given by
\begin{align}\label{eq.post.strat}
    \E[Y|R=0]= \sum_{x \in \mathcal{X}} \E[Y|X=x] p(X=x|R=0) = \sum_{x \in \mathcal{X}} \E[Y|X=x, R=1] p(X=x|R=0) 
\end{align}

\noindent Note that the post-stratification estimator is simply the empirical analog to the right-hand side of \eqref{eq.post.strat}. 

Unfortunately, as noted above, these non-parametric assumptions and procedures are insufficient in many practical cases because values of $X$ can appear in the target group that don't appear in the sample. This is certain to happen with continuous valued $X$, but also likely to happen even with discrete variables with more than a few dimensions and categorical values of discrete $X$. 
Turning to calibration estimators, we must import additional or different assumptions to achieve identification. We therefore rely on an assumption we term ``linear ignorability'',
\begin{assumption}[Linear ignorability in $\phi(X)$]\label{assumption.linearign}
Let $Y_i$ be written $\phi(X_i)^{\top}\beta + \epsilon_i$, and the probability of unit $i$ being sampled generated by $Pr(R_i=1) = g(\phi(X_i)^{\top}\theta + \eta_i)$, where $g(\cdot):\mathcal{R}\mapsto [0,1]$. Linear ignorability holds when $\epsilon_i \indep \eta_i$.
\end{assumption}

In words, this requires the part of $Y$ not linearly explainable by (i.e. orthogonal to) $\phi(X)$ to be independent of the part of the response process not linearly explained by $\phi(X)$ via a suitable link function. Under linear ignorability (with a given choice of $\phi(\cdot)$), a feasible calibration estimator using that choice of $\phi(\cdot)$ will be unbiased:
\begin{proposition}[Unbiasedness of Calibration under Linear Ignorability]\label{prop:unbiased1}
Under linear ignorability in $\phi(X)$ (Assumption~\ref{assumption.linearign}) the calibration estimator using weights chosen by Equation~\eqref{eqn:cal_estimator} will be unbiased for the target population mean of $Y$. 
\end{proposition}

\noindent Proof of Proposition~\ref{prop:unbiased1} can be found in Appendix~\ref{app:proofs}. Linear ignorability's connection to existing thinking can be found in the two well-known special cases that it covers, each of which is sufficient but neither of which is necessary to satisfy linear ignorability. At one extreme would be the assumption that $Y$ is truly linear in $X$ without unobserved confounders of $X$, meaning $\epsilon$ is fully independent of any other variable in the system, including the usual ``conditional independence assumption'',  $\E[\epsilon|X]$. It is important to note that to write ``$Y = \phi(X)^{\top} \beta + \epsilon $'' in Assumption~\ref{assumption.linearign} is not to require this, but only to invoke the decomposition of $Y$ into a component in the span of $\phi(X)$ and a residual piece $\epsilon$. Linear ignorability is slightly weaker as  it requires only that the $\epsilon$ formed by removing what is linearly-explainable by $\phi(X)$ is independent of $\eta$. That is, there can be unobserved confounders of $X$ and $Y$ here, which would appear as $\epsilon$ values that are not independent of $X$, but these are not problematic unless they are correlated with the unmodeled influences on selection, found in $\eta$.   

At the other extreme, one could assume \textit{link-linearity of the response model}, such that $\eta_i$ is independent of any variable in the system. Such assumptions are standard in prior work on calibration such as \cite{sarndal2005estimation, Kott:2010dq, zhao2016entropy}, though the choice of link functions in these is sometimes more restrictive than the general case we show here (see Appendix~\ref{app:proof_linear_outcome}). Linear ignorability is weaker than such an assumption, similarly, in that it requires only the part of $g^{-1}(Pr(R=1))$ not in the span of $\phi(X)$ to be independent of $\epsilon$. That is, there may be systematic influences on selection that go unmodeled and appear $\eta$, so long as these are unrelated to the unmodeled influences of $Y$, found in $\epsilon$.

\subsection{Bias due to violating linear ignorability}\label{subsec:counter}

To clarify the commitments one makes by subscribing to the linear ignorability assumption, we illustrate how it might be violated. Consider the decomposition of $Y$ as $\phi(X)^\top\beta + (Z + \nu)$. The $\epsilon$ invoked in Assumption~\ref{assumption.linearign} is composed here of $Z+\nu$. Here, $\nu$ is entirely exogenous random noise; 
$Z$ is unobserved and, without loss of generality, orthogonal to $\phi(X)$ because it could equivalently be replaced by the residual from projecting $Z$ onto $\phi(X)$. Whether linear ignorability holds is determined by $Z$'s role in the selection process. If $Z$ was purely exogenous random noise (like $\nu$) then $\epsilon = Z+\nu$ will be independent of $\eta$ in the equation for $R$, satisfying Assumption~\ref{assumption.linearign}. By contrast, if this $Z$ is associated with $R$ (and thus $\eta$, since it is independent of $X$), then $Z$ would cause a violation of Assumption~\ref{assumption.linearign}.

Problematic variables $Z$ could take on two forms. First, there could be important omitted variables, which would also violate non-parametric ignorability (Assumption~\ref{assumption.npignorable}).
Unobserved factors outside of $\phi(X)$ could be relevant to both $R$ and to $Y$, thus entering into both $\epsilon$ and $\eta$, causing them to be correlated. For example, an individual's general level of interest in politics is predictive of many policy positions, and the strength of those preferences, in American politics.  It is also highly predictive of response probability to political surveys, with those interested in politics over represented in respondent samples.  Because political interest is not measured in many datasets used to define target populations, such as those defined by administrative records, it is an example of an unmeasured confounder $Z$ that could violate both non-parametric ignorability and linear ignorability. No adjustment technique would eliminate bias in this scenario, but sensitivity analyses provide a natural approach to addressing potential remaining bias from such confounders \citep[e.g.][]{hartman_huang_2023}. We note that it is only the part of political interest orthogonal to the linear relationship with the included auxiliary variables in $\phi(X)$.  To the degree that the included auxiliary covariates also adjust for imbalance in political interest, and mitigate bias from an unobserved $Z$, it is only the remaining part of bias from such an omitted confounder that is of concern. 

The second form of problematic Z would be one that generates a ``specification failure''. Suppose we did not omit any variable ``important to'' $R$ and $Y$, but $Z$ is a nonlinear function of $\phi(X)$ (orthogonal to what is in $\phi(X)$), that is relevant to both $Y$ and $R$. $Z$ would then appear in both $\epsilon$ and $\eta$, driving their association.  This is of particular concern for the commonly used mean calibration in which $\phi(X) = X$. This form of $Z$ is difficult to rule out: investigators may suspect the outcome to ``involve'' an $X$ corresponding to some concept, but can rarely make strong arguments for the functional relationship to $R$ and $Y$, or justify a particular link function for $R$. Such a problematic form of $Z$ is the main motivation for our approach. Examples of how such a $Z$ emerges, the bias it generates, and the \kpop\ solution to it, are illustrated in first a simple reductive simulation and then a more complex one in Section~\ref{sec:simulations}.

\section{Proposal: kernel-based weighting (\kpop)}\label{sec:proposal}

Many reasonable proposals are possible for how to choose $\phi(X)$ so as to mitigate violations of linear ignorability and the consequent bias. In plain terms, we want $\phi(X)$ to capture any (potentially non-linear) systematic relationship between $Y$ and $X$ and/or $R$ and $X$. This would expunge problematic ``$Z$'' variables from $\epsilon$ and/or $\eta$, so that such a $Z$ can no longer drive an association of $\epsilon$ with $\eta$, thereby achieving linear ignorability. \kpop\ is designed to reduce the risk and magnitude of violating linear ignorability, with minimal user-intervention, by replacing the design matrix $\X$ with a kernel matrix $\K$ that represents a rich choice of $\phi(\cdot)$.

\subsection{Motivation for kernels through models}

One way to motivate the use of kernels is through considering how they determine the choice of $\phi(\cdot)$ in the context of linear models. Consider linear functions of $\phi(X)$ that explain either the outcome or the response probability (transformed as $g^{-1}(Pr(R=1))$) to be linear in $\phi(X)$. For simplicity we consider the outcome. Consider the notional regularized regression problem: 
\begin{align}\label{tikhonov}
    \argmin_{\theta \in \mathbb{R}^{P'}} \sum_i (Y_i - \phi(X_i)^{\top}\theta)^2 + \lambda \theta^{\top}\theta
\end{align}

This model will not actually be estimated in our setting---it is simply a problem statement that allows us to arrive at the kernel based approach to building basis functions, $\phi(X)$. Ideally our choice of $\phi(X)$ would be one that includes very general, high-dimensional, non-linear expansions of $X$. Fortunately, certain choices of $\phi(X)$ can be high- or infinite-dimensional, yet admit an $N_s$-dimensional representation of the data that can then be employed in calibration. 

Intuitively kernel functions ``compare'' two observations, $X_i$ and $X_j$ by computing $k(X_i,X_j):  \mathbb{R}^P \times \mathbb{R}^P  \mapsto \mathbb{R}$. A kernel function $k(\cdot,\cdot)$ is positive semi-definite if the kernel matrix it creates, $\K$, satisfies $a^{\top}\K a \geq 0$ for all real vectors $a$. For such positive semi-definite kernels, the value of $k(X_i,X_j)$ corresponds to choices of $\phi(X_i)$ through the relationships $k(X_i, X_j) = \langle \phi(X_i), \phi(X_j) \rangle$. As is well known and can readily be shown from first principles (see e.g. \citealp{krls}), the solution to Equation~\ref{tikhonov} admits to the form $\theta = \sum_i c_i \phi(X_i)$, and consequently the predictions for $Y_i$ are then given by $\phi(X_i)^{\top}\theta = \sum_j c_j k(X_i, X_j)$. Forming the kernel matrix $\K$ with entries $\K_{i,j} = k(X_i, X_j)$, this  can be rewritten as $K_i^{\top}c$ where $K_i$ is the $i^{th}$ row of $\K$, or the vector of predictions $\hat{\Y}$ is simply $\K c$. 

The vital feature of this result is simply that \textit{the functions linear in $\phi(X_i)$ have been replaced with those linear in $\K_i$}, or equivalently, the linear span of $\phi(X)$ is covered by the linear span of $\K$. This holds regardless of the dimensionality of $\phi(\cdot)$. Thus, to gain all the benefits of $\phi(\X)$ -- whether for modeling or calibration purposes -- one need only work with $\K$. 

Here we employ the Gaussian kernel, 
\begin{equation*}\label{eq:gauss}
  k(X_i,X_j) =  exp(-||X_i-X_j||^2/b)
\end{equation*}
\noindent where $||X_i-X_j||$ is the Euclidean distance. While no single choice of kernel can rigorously be established as optimal across settings or even in a particular application, the Gaussian kernel typically serves as the ``workhorse’' kernel for a wide variety of kernel based procedures. One reason for this is that the implicit $\phi(\cdot)$ for the Gaussian kernel is infinite-dimensional and has the ``universal representation property'', such that as the number of sample points goes to infinity, every continuous function will be linear in these features \citep{micchelli2006universal}. The values of $k(X_i,X_j)$ can readily be interpreted as a distance or similarity measure between $X_i$ and $X_j$, with $k(X_i, X_j)=1$ only when $X_i = X_j$, i.e. that the covariate profiles match exactly. The rate at which $k(X_i,X_j)$ approaches zero when $X_i$ and $X_j$ differ is dictated by the choice of $b$, which we discuss below. A linear combination of the elements of $K_i$ is thus a weighted sum of unit $i$'s similarity to every other unit $j$ in the sample, where similarity is measured by centering a Gaussian kernel over each $X_j$ and measuring its height at $X_i$.  \cite{krls} provides further description and illustration of this function space. 

\subsection{The ideal \kpop\ estimator}

In this section we describe the ``ideal'' \kpop\ estimator, which will be revised below to an approximate version. Replacing $\phi(X_i)$ with $K_i$,  we seek to satisfy Equation~\eqref{eq:balance} by choosing weights that achieve,
\begin{align}\label{eq:kweights}
  \sum_{i:R=1} w_i K_i = \frac{1}{\Np}\sum_{j=1}^{\Np} K_{j},\,\, s.t.\, \sum_i w_i=1,\,w_i \ge 0,\ \forall i
\end{align}

Note that every $K_i$ here is a transformation of $X_i$ that compares unit $i$ to each of the units in the survey sample. The matrix $\K$ has a row for every unit in the sample and in the target population, yielding dimensions $\Ns+\Np$ by $\Ns$.The term on the right gives an (unweighted) average row of $K_j$ for units in the target population. Note that each $K_j$ is an $\Ns$-vector, with the $i^{th}$ element indicating how similar unit $j$ in the target population is to unit $i$ in the survey sample, i.e. $k(X_j,X_i)$.  The term on the left is a weighted average of $K_i$ over the survey sample. Here too each $K_i$ is an $\Ns$-vector, with the $l^{th}$ element indicating how similar unit $i$ in the survey sample is to unit $l$ in the survey sample, i.e. $k(X_i,X_l)$.

In cases where (known) weights $w^{(pop)}$ are used to adjust the target population itself---as in our application below---then \kpop\ would instead seek weights that bring the weighted means of $K_i$ among the sampled units to approximately equal the $w^{(pop)}$-weighted means of $K_i$ in the target population. Thus the weighting condition in~\eqref{eq:kweights} becomes
\begin{align}\label{eq:kweightspop}
  \sum_{i:R=1} w_i K_i = \sum_{j = 1}^{\Np} w^{(pop)}_j K_{j},\,\, s.t.\, \sum_i w_i=1,\,w_i \ge 0,\ \forall i
\end{align}
\noindent This formulation is thus more general. We also include $w^{pop}$ in describing the bias bound and approximation routine below to accommodate cases where it is required.

Calibrating through this kernel transformation achieves balance on a wide range of non-linear functions of $X$, without requiring the researcher to pre-specify them. For example, as we will show in Section \ref{sec:application}, $\kpop$ achieves balance on the interaction of education level, region, and race in a 2016 U.S. Presidential survey without requiring the researcher to have foreknowledge of its importance, much less requiring specific knowledge that Midwestern, white voters with lower levels of educational attainment must be accounted for in the survey weights to yield accurate national predictions.

Another view of what these weights achieved, discussed in \cite{hazlett2020kernel},  is that approximate balance on a kernel transformation approximately equates the multivariate distribution of $X$ in the two groups, \textit{as it would be estimated by a corresponding kernel density estimator}. We also note closely related work on kernel-based balancing and imbalance metrics including \cite{wong2018kernel, yeying2018kernel, kallus2020generalized}. \cite{tarr2021estimating} consider a related approach implemented by interpreting the Lagrange coefficients estimated in a support vector machine with such a kernel as weights. 

\subsection{The necessity of approximate balance}\label{sec:approxbalance}

Weights that achieve equal means on every element of $K_i$, i.e. every column of $\K$, are often infeasible. Even where this can be achieved within numerical tolerance, such calibration could lead to extreme weights. Instead, we use approximate calibration weights designed to minimize the worst-case bias due to remaining miscalibration. While numerous approximation approaches are possible, we use a spectral approximation.  Specifically, $\K$ has singular value decomposition $\K = \V \A \U^{\top}$, where the columns of $\V$ are left-singular vectors. Even granting that the linear ignorability assumption holds, approximate balance means the calibration step is not complete, which can introduce \textit{additional} bias, referred to here as the approximation bias. The worst-case bound on this approximation bias is given by \cite{hazlett2020kernel}
\begin{align}\label{eq:worstcase}
\sqrt{\gamma} ||(w^{(pop) \top}\V_{pop} - w_s^{\top}\V_s) \A^{1/2}||_2
\end{align}
where $\V_{pop}$ is the matrix containing the rows of $\V$ corresponding to target population units, $\V_s$ contains the rows of $\V$ corresponding to sampled units, and $\A$ is the diagonal matrix of singular values. In this bias bound, $w^{(pop)}$ denotes the (optional) known weights for adjusting the target population itself. The scalar $\gamma$ is the (reproducing kernel Hilbert) norm on the function, equal to $c^{\top}\mathbf{K}c$ effectively describing how complicated or ``wiggly'' the chosen function is.  This is an unknown constant that need not be estimated during the optimization we describe below.

We make three remarks on the form of this worst case approximation bias in Equation \eqref{eq:worstcase}. First, the $L_2$ norm of the regression function ($\sqrt{\gamma}$) controls the overall scale of potential bias. Second, the imbalance on the left singular vectors of $\mathbf{K}$ after weighting, $(w^{(pop) \top}\V_{pop} - w_s^{\top}\V_s)$, enters directly. Third, the impact of imbalance on each singular vector is scaled by the square root of the corresponding singular value.

The third point in particular suggests the approximate balancing approach we use: calibrate to obtain nearly exact balance on the first $r$ singular vectors (columns of $\V$), leaving the remaining ($r+1$ to $\Ns$) columns uncalibrated. The choice of $r$ is then chosen to minimize the bias bound (Equation \eqref{eq:worstcase}). In practice, the singular values of a typical matrix $\K$ decrease very rapidly (see Appendix \ref{app:scree_k} for an illustration from the application below). Thus, balance on relatively few singular vectors achieves much of the goal, though the procedure continues beyond this to minimize the worst-case bias bound in Equation~\ref{eq:worstcase} directly.
 
\paragraph*{``Mean-first'' \kpop\: A ``no-worse'' solution.}\label{sec:mf_routine} Achieving approximate balance on $\K$ will typically yield good, but not perfect balance on the means of the original variables, $\X$. In practice a visible difference in means or proportions on a given variable can be unsettling: researchers and pollsters may reasonably hope for nearly exact mean calibration on variables of known importance to the outcome of interest, even if the means are in fact no more important to balance than unseen higher-order moments. Further, it may be useful to know that a given estimator achieves balance on the same moments as conventional raking or mean calibration, in addition to possibly calibrating higher-order moments.

To this end,  we advocate for using a ``mean first'' procedure, in which the weights are constrained to obtain equal means (within a set tolerance) on a chosen set of variables $\X$, in addition to calibrating on $r$ singular vectors of $\K$ chosen so as to minimize the bias bound described above. The cost of enforcing mean balance is that there may be fewer dimensions of $\K$ that can additionally be balanced on within feasibility constraints. Nevertheless, the virtue of this approach---at least as a transitional methodology---is that in terms of balance and anticipated bias it is arguably ``no worse'' than the conventional approach of calibrating on the means of $\X$ alone. For improved stability and performance in practice, we recommend an approximate balancing approach that appends the left-singular vectors of the chosen set of $\X$, choosing the number to balance on by minimizing the worst case bias bound in Equation \eqref{eq:worstcase}. We refer to this as \textit{kpop+mf} below.

\paragraph*{Inference.}\label{app:SEs} Following the calibration weighting literature, we use a linearized variance estimator \citep{fuller1975regression, kott2016}.  Due to the often large number of dimensions of $\mathbf{K}$ chosen by the method described in Section~\ref{sec:approxbalance} for the \kpop\ calibration constraints, we use a ridge regularized regression of the outcome on the $r$ selected columns of $\mathbf{K}$, leaving any columns corresponding to ``mean first'' constraints unregularized if they are included.  In Appendix \ref{sims:SEs} we show performance of these standard errors, which accurately estimate the empirical standard error and achieve near nominal coverage rates in our simulations.

\subsection{Choice of kernel, data scaling, and \emph{b}}\label{sec:kernel_specifics}

One obstacle to adopting kernel-based methods is that while they can greatly reduce researcher-degrees-of-freedom in terms of selecting $\phi(\cdot)$, they do still require choosing the kernel function and the value of any of its hyperparameters. In this work we employ the Gaussian kernel, which we regard as reasonable on account of its universal representation property \citep{micchelli2006universal}. There are important considerations regarding how $\X$ is scaled, and relatedly, the choice of the Gaussian kernel bandwidth, $b$.

\paragraph{Data scaling.}Prior to constructing the kernel, continuous-valued $X$ are scaled to have a variance of one. Such a choice is convenient as it ensures no ``unit of measure'' choice will affect the results. Under this standardization, a one-standard-deviation difference on a given continuous covariate will add one to the squared Euclidean distance that forms the numerator of the exponents in the kernel function. 
For categorical variables, there is no added distance between two observations that have the same value of a given variable. However, if two units have different values on a categorical variable, we scale the data such that it adds a distance of one to the numerator of the exponent in the kernel function. This is one choice that keeps categorical and continuous variables on reasonable relative scales in terms of their influence in the kernel function. We can achieve this scaling numerically simply by (i) one-hot encoding all binary and categorical variables (without dropping a level), (ii) rescaling those one-hot encoded indicators by $1/\sqrt{2}$.

\paragraph{Kernel sampling and feasibility}

In the present setting we rely only on the observation in the sample to formulate the columns of a kernel matrix. This is because (i) if there are millions of observations in the target population, constructing a matrix with that many rows would be infeasible, and (ii) the representation of each unit based on its similarly to other units in the sample is most relevant to how we reweight members of the sample; if there were members of the population that are very different from members of the sample, then no weighting of the sample will account for this.  

\paragraph{Kernel bandwidth.}The choice of $b$ in the kernel definition scales the similarity measure and is thus effectively a feature extraction choice, constrained by feasibility. Too small a choice of $b$ makes each observation appear ``too unique,'' pushing the kernel distance, $exp(-||X_i - X_j||^2/b)$, to zero for any given pair of units; on the other hand, too large a choice of $b$ makes each observation seem ``too similar,'' producing a kernel distance approaching one for all pairs. A choice of $b$ is therefore desirable when it yields a $\K$ with meaningful variability in the similarity measure among different pairs of units. We use the variance of $\K$ as a measure of the useful information available at a given choice of $b$ and turn to this metric to motivate our choice of bandwidth, selecting the value which produces maximal variance in $\K$. We make no claim as to the optimality of this result, but it offers a reasonable choice that can be established without looking at the result. In our simulations and applications, this choice produces consistently good performance, though the results are shown to be stable across a wide range of $b$ regardless (see appendix \ref{app:various_b}). Further discussion of the kernel bandwidth, as well as details on data pre-processing and scaling decisions appropriate for categorical, continuous, or mixed variable settings can be found in Appendix~\ref{app:kernelchoice}.

\subsection{Practice and diagnostics}\label{sec:kernel_diag}

We recommend several diagnostics that can be used to better understand the resulting weights and what they achieve or fail to achieve. First, the number of dimensions of $\K$ optimally selected for calibration ($r$) should be checked. If this is very small (e.g. 1 or 2), the user should be aware that little balance was achieved.  Next, researchers should compare the weighted sample and target population margins on the original $\X$ and explicitly chosen functions of these variables that may be of concern, such as interactions. We illustrate this below. Third, we suggest two summary statistics to assess the degree to which multivariate balance has been improved. The first is an $L_1$ measure of the distance between the distribution of $X$ for the survey and the population, summed over the units of the survey. This can be obtained both before and after weights are applied to assess the reduction in multivariate imbalance \citep{hazlett2020kernel}. The second is the ratio of the bias bound  (Equation~\eqref{eq:worstcase}), calculated with and without the weights, to determine the proportional improvement in the degree of potential bias due to remaining imbalances on $\K$. Both serve to indicate to the user whether substantial improvements in multivariate balance were achieved by the weights. 

Finally, it is often valuable to understand how extreme the weights are and thus how heavily the solution depends on a small number of observations. This can be done by the investigator's preferred means, such as inspecting the distribution of weights visually, or constructing statistics such as the effective sample size or the number of observations (working from the most heavily weighted towards the least) that one needs to sum to achieve 90\% of the total sum of weights. We present these diagnostics for our application in \ref{app:dist_w}.

\section{Simulations}\label{sec:simulations}

\subsection{A minimal simulation example}\label{sec:minimal_example}

We provide here an example to illustrate how easily bias can emerge in a simple case with just two variables and with mean balance holding perfectly by construction. Suppose a target population of interest consists of four groups in equal shares: college-educated females, college-educated non-females, non-college-educated females, and non-college-educated non-females. A given policy happens to be supported by 80\% of college-educated females and only 20\% of those in the other three groups. Thus, the mean level of support in the target population would be 0.25(0.8)+0.75(0.2) = 35\%. A sample is designed to carefully quota on gender and education, obtaining 50\% female and 50\% college-educated respondents. We use quota sampling in this example for simplicity as it allows us to have a sample already matched to the target population on the margins. The same considerations would apply, however, in a convenience sample or more generally if weighting were required to achieve mean calibration.

However, this sampling process neglects the proportion of each intersectional stratum. Suppose that, among females, the sample drew a higher proportion of college-educated respondents (three-quarters, as opposed to half in the target population). Conversely among non-females, suppose that fewer respondents were college-educated (one-quarter, instead of half). The average level of support for this policy in the unweighted sample would then be $\frac{1}{2}\frac{3}{4}(0.8) + \frac{1}{2}\frac{5}{4}(0.2)= 42.5\%$, rather than the 35\% in the target population.  In other words, a key interaction term (female$\times$college), or the indicator for being in that intersectional stratum, has a different mean in the sample and the target population, and it influences the outcome. It is thus an example of an omitted variable, $Z$, that drives an association between $\epsilon$ and $\eta$, violating Assumption~\ref{assumption.linearign} unless $Z$ was included. Table~\ref{tab:mean_cal} summarizes this situation, and describes a set of ``ideal weights'' that would correct the proportions of each intersectional stratum, thus producing the correct answer. 

\begin{table}[ht!]
\centering
\resizebox{\textwidth}{!}{%
\begin{tabular}{cc|cc|c|ccc}
   \multicolumn{2}{c}{characteristics} &  \multicolumn{2}{|c|}{proportions} & outcome & \multicolumn{3}{c}{weights (times $\Ns$)} \\
    \hline
  female & college & target population & sample & Pr(support) & unweighted & mean cal. & ideal \\
  \hline
   1 & 1 & 1/4 & 3/8 & 0.80 & 1 & 1 & 2/3 \\ 
   1 & 0 & 1/4 & 1/8 & 0.20 & 1 & 1 & 2 \\ 
   0 & 0 & 1/4 & 3/8 & 0.20 & 1 & 1 & 2/3 \\ 
   0 & 1 & 1/4 & 1/8 & 0.20 & 1 & 1 & 2 \\ 
   \bottomrule
   \multicolumn{4}{r}{target population mean:} & \multicolumn{1}{c}{0.35} & & & \\
   \multicolumn{4}{r}{weighted mean:} & \multicolumn{1}{c}{} & \multicolumn{1}{c}{0.425} & 0.425 & 0.35
\end{tabular}}
\caption{\label{tab:mean_cal} Quota sampling ensured the sample was representative on the means of college and female. Mean calibration weights can thus be uniform. College-educated females respondents are over-represented  in the sample, as are non-college educated non-females, leading to mean calibration failing to balance all strata of $X$. The ``ideal'' weights represent the choice that would bring the sample proportion of each stratum to match that in the target population.
}
\end{table}

The mean calibration weights indicated in Table~\ref{tab:mean_cal} are uniform because the quota sampling already ensured mean balance on each characteristic. The ideal weights are those that would bring each stratum's sample proportion to match its target proportion. For example, the female-college category is 3/8 of the observed sample, but should be brought to 1/4. Weighting by 2/3 achieves this, as (3/8)(2/3)=1/4. Note that in this simple setting with just two binary variables, this would be feasibly and perfectly achieved by using post-stratification. The challenge will be in more complex examples, where post-stratification is infeasible, as shown in our next example and application.  

Turning now to the application of \kpop\ in this example, Table~\ref{tab:k.example} presents one corner of the kernel matrix $\K$ for each of the four unique unit types. For any two units $i$ and $j$ with the same values on the covariates, $k(X_i,X_j)=e^{0}=1$. Thus, the diagonal of the kernel matrix will always be one. Choosing $b$ conveniently for illustrative purposes, individuals that differ on one trait but not the other will have $k(X_i,X_j)= e^{-\left((1-0)^2 + (0-0)^2 \right)} = e^{-1} \approx 0.37$. Individuals who differ on both characteristics will have $k(X_i, X_j)= e^{-\left((1-0)^2 + (1-0)^2\right)}= e^{-2} \approx 0.14$.

\begin{table}[!ht]
    \centering
    \resizebox{\textwidth}{!}{%
    \begin{tabular}{cc|c|c|ccccc|cc}
     \multicolumn{2}{c}{characteristics ($\X$)} &  & \multicolumn{6}{|c|}{kernel matrix ($\K$)} & outcome & weights \\ \hline 
    female & college & sample \% & & k(,1) & k(,2) & k(,3) & k(,4) & (repeats) & Pr(support) & (kpop) \\ \hline
     1 & 1 & 3/8 & k(1,) & 1 & .37 & .14 & .37 &  \ldots & 0.80 & 2/3 \\
     1 & 0 & 1/8 & k(2,) & .37 & 1 & .37 & .14 & \ldots & 0.20 & 2\\
     0 &  0 & 3/8 & k(3,) & .14 & .37 & 1 & .37 &  \ldots & 0.20 & 2/3 \\ 
     0 & 1  & 1/8 & k(4,) & .37 & .14 & .37 & 1 &  \ldots & 0.20 & 2\\
     & &  & \vdots & \vdots & \vdots & \vdots & \vdots & \vdots &  & \\ \hline
    \multicolumn{2}{r|}{target mean:} & & &  0.47 & 0.47 & 0.47 & 0.47 & & 0.35 & \\
    \multicolumn{2}{r|}{sample mean:} & & & 0.52 & 0.42 & 0.52 & 0.42 & & 0.425 & \\
    \end{tabular}%
    }
    \caption{Kernel matrix representing each of four unique types of individuals in the sample. Each element $k(X_i, X_j)$ is equal to $exp(-||X_i-X_j||^2/1)$, where the numerator in the exponent will be equal to two times the number of features on which $i$ and $j$ differ. The columns provide new bases for representing the data.
    }
    \label{tab:k.example}
\end{table}

Since there are equal numbers of individuals of each type in the target population, the average observation is equally ``similar to'' individuals of all types -- the target means for each column of $K$ is the same. By contrast, in the sample, the over-representation of the first and third categories (three of eight units in female$\times$college, three of eight units in non-female$\times$no college) leads the average observation to be overly ``similar to'' observations of these over-represented types and insufficiently ``similar to'' the under-represented types. \kpop\ will seek weights for the sample units that make the (weighted) average columns of $K$ over the sample equal the average in the target population. The weights that achieve this are shown in the final column. These will have the effect of downweighting units in the female$\times$college and non-female$\times$no college categories (with a weight of 2/3), and upweighting the remaining two categories (with a weight of 2). This reproduces exactly the ``ideal'' weights in Table~\ref{tab:mean_cal}. Consequently, with only the matrix $\X$ for the sample and target population and no further information about the importance of the interaction, the \kpop\ weighted estimate of the mean level of support matches that in the target population (35\%).
 
\subsection{Realistic simulation setting} \label{subsec:sim_intro}

The above example was made as simple as possible for purposes of clarifying the method. In practice, however, methods such as post-stratification would have also worked in that setting.  We next consider a more complicated setting to demonstrate, first, the approach in a context where other methods will encounter difficulties, and, second, to more fully illustrate performance on both bias and variability. We design this simulation to closely match the application below, but in this case we specify the selection and outcome models within the simulation. We first construct a model through which respondents in the smaller sample are drawn from the larger target population. In our case, the target population is given by the post-election wave of the 2016 Congressional Cooperative Election Study (CCES) \citep{DVN/GDF6Z0_2017}. We specify our selection model $p(S=1) = logit^{-1}\left( \mathbf{X}\beta\right)$ and construct new samples by taking Bernoulli draws from the CCES population according to this model. Our outcome model is linear in the same set of covariates, $p(Vote=Dem) = \mathbf{X}\gamma$, allowing us to directly control the mechanism of bias by specifying the correlation of $\beta$ and $\gamma$. In the following simulation, this correlation is about -.75, producing negative bias of about -3.5 p.p. in the unweighted sample. 

Both models are (link) linear in the same, fairly simplistic set of covariates $\X$: party identification, age (4-way), gender, education (3-way), race (4-way), born-again Christian status, and a subset of two-way interactions between party-id and age as well as born-again status and age. Coefficients in the selection model are chosen to produce roughly realistic samples comparable to the observed Pew survey and scaled to produce a sample size of roughly 500. For the outcome model, coefficients are adjusted through an automated procedure to produce values in probability scale.   Below, we present results across 1000 iterations. Further details can be found in Appendix ~\ref{app:simulations}.

\begin{table}[!ht]
\caption{Simulation Results}
\begin{centering}
\begin{tabular}[t]{lrrr}
\toprule
  & Bias (p.p.) & MSE & Abs Bias Reduction (\%)\\
\midrule
unweighted & -3.510 & 12.603 & 0.000\\
mean calibration (demos) & -1.618 & 2.893 & 0.539\\
mean calibration (demos+edu) & -1.296 & 1.961 & 0.631\\
mean calibration (all) & -0.029 & 0.226 & 0.992\\
kpop & -0.272 & 0.357 & 0.923\\
kpop+mf (demos) & -0.165 & 0.297 & 0.953\\
kpop+mf (demos+edu) & -0.150 & 0.268 & 0.957\\
kpop+mf (all) & 0.012 & 0.244 & 0.997\\
Horvitz-Thompson (true) & -0.160 & 10.229 & 0.954\\
post-stratification (true) & -1.120 & 1.571 & 0.681\\
mean calibration (true) & -0.010 & 0.214 & 0.997\\
\bottomrule
\end{tabular}
\caption{Bias, mean squared error, and absolute bias reduction by weighting method across 1000 simulations wherein the outcome and selection model are specified using the same set of variables to directly control the mechanism of bias. The models above the line represent specifications that investigators might realistically attempt without  access to the unknowable, true selection model.
For comparison, those below the line demonstrate the performance of estimators given ``true'' information about the correct selection model that would be unknown to investigators.
}
\end{centering}
\end{table}

We compare several \kpop\ and \textit{kpop+mf} specifications to a range of approaches that we anticipate thoughtful investigators might attempt, as well as three approaches that exploit the true specification or selection probability for comparison. These include raking on just the basic demographic variables (\textit{mean calibration (demos)}), on demographics and education (\textit{mean calibration (demos+edu}), and on all available variables (\textit{mean calibration (all)}). See Appendix \ref{tab:margins_table} for additional information on these variables. We motivate these specifications and why investigators might choose them in the context of the application (Section \ref{sec:application}). 

Finally, for benchmarking purposes we compare \kpop\ as it would be implemented (without access to true information about the model or variables to include) with three methods that do have access to this information: (i) post-stratification on the correct (minimal) set of intersectional variables described (\textit{post-stratification (true)}); (ii) the Horvitz-Thompson estimator employing the true (unknown) sampling probability (\textit{Horvitz-Thompson (true)}); and (iii) mean calibration on just the variables required for linear ignorability (\textit{mean calibration (true)}).

We find that the four \kpop\ specifications all significantly reduce the bias (by 92-99.9\%) and have MSEs roughly an order of magnitude smaller than those of \textit{mean calibration (demos)} and \textit{mean calibration (demos+edu)}, both of which reduce bias only by about half. By contrast  \textit{mean calibration (all)} happens to perform very well, highlighting the specification sensitivity of this approach as compared to the stability of \kpop.  Further, \kpop\ performs well even compared to estimators that are given access to the true model or set of variables to include. Even with our fairly simplistic selection model, \textit{Post-stratification (true)} still struggles with  the ``empty cell'' problem, dropping on average 23\% of population units and reducing bias only by 68\%. Notably, the Horvitz-Thompson estimator is roughly unbiased as expected, but has an MSE almost as large as the unweighted estimator and over 30 times larger than any \kpop\ estimator. \textit{Mean calibration (true)} performs well here, but is still comparable to the more naive \kpop\ estimators. 

\section{Application: 2016 U.S. Presidential Election}\label{sec:application}
In the 2016 United States Presidential election, state-level polls in key states were severely biased, with polling aggregators making over-confident predictions that Donald Trump would lose. National polls correctly predicted that Hillary Clinton would lead the national popular vote, while many overstated the margin. The challenges of correctly weighting a highly non-random sample to match the national electorate likely contributed to these errors. As \citet{kennedy2018evaluation} note, existing polls were especially likely to over-represent college-educated whites. 

We test whether weighting with \kpop\ would have improved on this, absent foreknowledge of what functions of covariates and intersectional strata are essential to address sources of bias. Because voters may have changed their mind between a given pre-election survey and the day of their vote, simply checking whether weighting the outcome of a pre-election survey produces an estimate close to the true election result would not provide a meaningful test of weighting techniques.  We instead estimate what the average ``retrospective vote choice'', measured post-election, would have been among voters in the 2016 election.  This involves (1) training a model that predicts stated retrospective vote choice as a function of $X$ using a large post-election survey which we define as the target population; (2) applying this model to predict the ``retrospective vote choice'' of each individual in a pre-election survey using their covariates $X$; (3) constructing weights to calibrate the pre-election sample to the target population; then (4) comparing the weighted average of the predicted ``retrospective vote choice'' in the pre-election sample to the stated vote choice the target population.  We emphasize that, were it not for the dynamic nature of vote intention in pre-election surveys, using this modeling outcome would not be necessary in settings where we can directly estimate the outcome, $Y$, of interest.

\subsection{Data and details}\label{subsec:app_data_details}
\paragraph{Survey sample.} For the respondent sample, we use survey data from the final poll conducted by the Pew Research Center before the general election in November 2016. Pew is a high-quality, non-partisan public opinion firm. The survey was conducted from October 20-25, 2016 using telephone interviews among a national sample of 2,583 of adults. On landline phone numbers, the interviewer asked for the youngest adult currently home (647), and cell phone interviews (1,936) were done with the adult who answered the phone \citep{pew2016}. Random-digit dialing was used, combined with a self-reported voter registration screen.  We keep only the $\Ns=2,052$ respondents who report that they plan to vote or have already voted. The publicly available data do not include survey design weights, and we use $q_i = 1$ for all respondents, although researchers could let $q$ be defined using design weights or previous calibration weights.  The survey includes proprietary multistage calibration weights, where the first-stage accounts for differential sampling probabilities due to the random-digit-dialing procedure, and the second-stage conducts a calibration procedure to match the U.S. population on many of the same auxiliary variables as we include.  We do not include these weights these weights as our base weights because we are weighting to a different target population, namely one defined by verified voters from a national survey, the CCES (discussed below), and under Assumption~\eqref{assumption.linearign}, our estimator is unbiased even starting from equal weights. Finally, we code vote choice as being for ``Republican Donald Trump'', ``Democrat Hillary Clinton'', or ``Other/Don't Know'', and we include voters who ``lean'' towards one of the two major party candidates.

\paragraph{Defining the target population.} Ideally we would define the target population using verified voter records from the Secretaries of State.  However, we do not have access to such an administrative file. Instead, following \citet{caughey_elements_2020}, we define our target population using the common content from the post-election wave of the 2016 Congressional Cooperative Election Study (CCES) \citep{DVN/GDF6Z0_2017}. The CCES is a large survey that aims to be representative of all voters, and the survey weights for the post-election wave lead to an estimate of the popular vote margin between the two major parties (2.48 percentage points, D - R) that is very close to the truth (2.3 percentage points).  Second, the CCES includes a number of demographic survey questions that overlap with those asked in the Pew study which we can use for calibration. We incorporate the weights provided by the CCES (``commonweight\_vv\_post'') into the definition of our target population. Limiting the data to voters for whom the outcome variable was not missing, and who stated that they ``definitely voted'' leaves a total of $\Np = 44,932$ units.

Our auxiliary data, $\X$, are defined using all of the overlapping variables in our data sets: age, reported gender, race/ethnicity, geographic region, education level, party identification, income, born-again Christian status, church attendance, and religion. All variables are self-reported except for region.  Appendix Table~\ref{tab:margins_table} summarizes the distributions of these variables and how they differ in the target population (CCES) compared to the survey sample (Pew).  For example, those with higher levels of educational attainment and higher income are over-represented in the Pew sample, as are older voters and Independents. By contrast Black voters and women are under-represented in the sample relative to the target population. 

\paragraph{Modeled outcome.} As noted above, the outcome variable to be weighted in this example is itself a modeled quantity representing the difference in probability of voting Democratic vs. Republican given one's covariates ($p(D_i-R_i|X_i)$). We estimate this using a multinomial logit model (see e.g. \citealp{long1997regression})
to estimate the relationship between $\X$ and three-way ``retrospective vote choice''  (Republican, Democrat, and Other) measured by asking respondents who they voted for in the post-election CCES survey. We use regularization in doing so, to mitigate over-fitting concerns (see e.g. \citealp{hastie2009elements}). This model includes gender, 3-way party identification, race/ethnicity, 6-way education, region, 6-way income, 5-way religion, 4-way church attendance, born-again status, continuous age, age$^2$, gender $\times$ party identification, and age (continuous) $\times$ party identification.

Recall that our goal is to find weights for the Pew observations (sample) such that the weighted average value of predicted $p(D_i-R_i|X_i)$ matches that in the CCES data, here 2.48 percentage points. None of the subsequent weighting methods are aware this particular choice of $\phi(X)$ has been made. Using this specification, the outcome can be modeled quite effectively. For example, choosing the highest-probability outcome as an individual's final vote choice leads to an 85-86\% correct classification rate for non-independents. This fitted post-election outcome model is then used to predict $p(D_i-R_i|X_i)$ using the $\X$ data from the Pew pre-election survey. Additional details can be found in Appendix~\ref{app:outcome_model}.

\paragraph{Weighting methods.} \label{subsec:appmethods}
We compare \kpop\ estimators to the two common methods researchers use for constructing survey weights discussed above, mean calibration and post-stratification. For mean calibration, we consider four specifications that represent a range of choices thoughtful researchers might attempt: (i) basic demographic variables only \textit{mean calibration (demos)}, including: age (4way), gender, race/ethnicity, geographic region, and party identification; (ii) those variables plus education information (\textit{mean calibration, (demos + edu)}); all available data (\textit{mean calibration, (all)}), adding  adding income, religion, born-again Christian status, and church attendance (see Table \ref{tab:margins_table}). Finally, we include one model that is given retrospective benefit, based on the analysis of \citet{kennedy2018evaluation}:  \textit{mean calibration (retrospective)}, which includes the interaction of party identification with age (4way) and, separately, with gender. It additionally addresses the importance of low-education white voters in the 2016 election, particularly in Midwestern states, by also including party identification $\times$ region $\times$ race/ethnicity for all voters. Among white respondents, this is expanded to also interact with education level (6way). We include this model  to evaluate the question of whether our proposed \kpop\ method can perform as well as a retrospective-informed model that serves as a best-case for what expert knowledge could hope to achieve.

Turing to post-stratification, using all 10 available variables results in highly complex cross-sectional strata which, in turn, make missing cells a significant hurdle when reweighting the survey sample, with nearly 92\% of population units representing strata not present in the sample. To bring the number of empty cells to a reasonable level, we coarsen age and income into 3- and 4-category variables respectively and do not include religion or church attendance. In order to give post-stratification the best advantage possible, we omit full 6-way education and instead stratify on a coarsened version of the above mentioned important, retrospectively-informed interaction among race and education, namely, coarsened 3-way education among white with no stratification by education among non-white voters. The resulting estimator stratifies on gender, race, region, party identification, born-again status, 3-way income, 3-way age, and the 3-way education and white interaction. Unfortunately, this still results in dropping around 30\% of units due to empty strata.

We compare the models described above against \kpop, applying our proposed kernel-balancing method with all the available categorical variables available as described in Table \ref{tab:margins_table}. For comparison with the preceding raking specifications, we also include three models, \textit{kpop + mf (demos), kpop + mf (demos + edu)}, and \textit{kpop + mf (all)} that first conduct mean calibration before proceeding to balance on the kernel matrix, as discussed in section \ref{sec:mf_routine}. 

\subsection{Results}

\paragraph{Balance.} We first consider the balance achieved by each method on the observed covariates. Table~\ref{tab:abs_error} presents the absolute error, weighted by the target population proportion for each level, for each auxiliary variable (rows 1-10) and a set of interactions (11-17). By construction, the mean calibration methods, as well as the \kpop\ \textit{+ mf} methods, perfectly match the marginal distributions for any variables that are included in the model.  

All methods greatly improve representativeness in the respondent sample as indicated by the reduction in error across variables and interactions relative to the unweighted sample. As expected, \kpop\ (without ``mean first'') achieves good but imperfect balance on the included covariates and interactions, despite not being directly constrained to achieve balance on them. Though we should expect post-stratification to produce perfect balance on the included terms, we see that, even with significant variable coarsening, empty cells pose a significant problem. As a result, post-stratification fails to get the correct margins, much less produce the non-parametric adjustment one hopes for under Assumption \ref{assumption.npignorable}.

\begin{table}

\caption{Weighted Mean Absolute Error on Auxiliary Variables (percentage points) \label{tab:abs_error}}
\begin{center}
\resizebox{\textwidth}{!}{%
\begin{tabular}[t]{lcc>{}c>{}c>{}c>{}c>{}c>{}c>{}c>{}c}
\toprule
  & \makecell[l]{Pew\\ Orig} & kpop & \makecell[c]{kpop+mf\\(demos)} & \makecell[c]{mean calib\\(demos)} & \makecell[c]{kpop+mf\\(d+edu)} & \makecell[c]{Mean Calib\\(d+edu)} & \makecell[c]{kpop+mf\\(all)} & \makecell[c]{Mean Calib\\(all)} & \makecell[c]{Mean Calib\\(retro)} & \makecell[c]{post-strat\\(reduc)}\\
\midrule
female & 3.65 & 0.11 & \textcolor{gray}{0.00} & \textcolor{gray}{0.00} & \textcolor{gray}{0.00} & \textcolor{gray}{0.00} & \textcolor{gray}{0.00} & \textcolor{gray}{0.00} & \textcolor{gray}{0.00} & \textcolor{gray}{0.06}\\
pid (3way) & 2.53 & 0.28 & \textcolor{gray}{0.00} & \textcolor{gray}{0.00} & \textcolor{gray}{0.00} & \textcolor{gray}{0.00} & \textcolor{gray}{0.00} & \textcolor{gray}{0.00} & \textcolor{gray}{0.00} & \textcolor{gray}{0.45}\\
age (4way) & 4.85 & 0.22 & \textcolor{gray}{0.00} & \textcolor{gray}{0.00} & \textcolor{gray}{0.00} & \textcolor{gray}{0.00} & \textcolor{gray}{0.00} & \textcolor{gray}{0.00} & \textcolor{gray}{0.00} & \textcolor{black}{1.95}\\
race/ethnicity (4way) & 1.54 & 0.16 & \textcolor{gray}{0.00} & \textcolor{gray}{0.00} & \textcolor{gray}{0.00} & \textcolor{gray}{0.00} & \textcolor{gray}{0.00} & \textcolor{gray}{0.00} & \textcolor{black}{1.25} & \textcolor{gray}{3.50}\\
region (4way) & 1.50 & 0.02 & \textcolor{gray}{0.00} & \textcolor{gray}{0.00} & \textcolor{gray}{0.00} & \textcolor{gray}{0.00} & \textcolor{gray}{0.00} & \textcolor{gray}{0.00} & \textcolor{black}{0.58} & \textcolor{gray}{2.51}\\
educ (6way) & 8.64 & 0.32 & \textcolor{black}{0.12} & \textcolor{black}{8.68} & \textcolor{gray}{0.00} & \textcolor{gray}{0.00} & \textcolor{gray}{0.00} & \textcolor{gray}{0.00} & \textcolor{black}{1.67} & \textcolor{black}{2.09}\\
income (6way) & 4.35 & 0.35 & \textcolor{black}{0.21} & \textcolor{black}{4.04} & \textcolor{black}{0.15} & \textcolor{black}{3.35} & \textcolor{gray}{0.00} & \textcolor{gray}{0.00} & \textcolor{black}{3.12} & \textcolor{black}{1.78}\\
born-again (bin) & 1.72 & 0.14 & \textcolor{black}{0.01} & \textcolor{black}{2.95} & \textcolor{black}{0.10} & \textcolor{black}{0.18} & \textcolor{gray}{0.00} & \textcolor{gray}{0.00} & \textcolor{black}{0.62} & \textcolor{gray}{4.95}\\
religion (5way) & 6.42 & 0.48 & \textcolor{black}{0.34} & \textcolor{black}{5.24} & \textcolor{black}{0.19} & \textcolor{black}{6.66} & \textcolor{gray}{0.00} & \textcolor{gray}{0.00} & \textcolor{black}{7.77} & \textcolor{black}{6.17}\\
church attnd. (4way) & 12.88 & 0.44 & \textcolor{black}{0.17} & \textcolor{black}{12.13} & \textcolor{black}{0.16} & \textcolor{black}{11.92} & \textcolor{gray}{0.00} & \textcolor{gray}{0.00} & \textcolor{black}{12.61} & \textcolor{black}{12.11}\\
pid$\times$race & 0.71 & 0.50 & \textcolor{black}{0.34} & \textcolor{black}{0.86} & \textcolor{black}{0.15} & \textcolor{black}{0.68} & \textcolor{black}{0.63} & \textcolor{black}{1.15} & \textcolor{black}{0.42} & \textcolor{black}{1.20}\\
educ$\times$pid & 6.22 & 0.36 & \textcolor{black}{0.21} & \textcolor{black}{6.25} & \textcolor{black}{0.29} & \textcolor{black}{0.45} & \textcolor{black}{0.39} & \textcolor{black}{0.57} & \textcolor{black}{1.56} & \textcolor{black}{1.49}\\
educ$\times$pid$\times$race & 3.81 & 0.45 & \textcolor{black}{0.48} & \textcolor{black}{3.87} & \textcolor{black}{0.34} & \textcolor{black}{0.44} & \textcolor{black}{0.67} & \textcolor{black}{0.58} & \textcolor{black}{0.46} & \textcolor{black}{0.90}\\
race$\times$educ$\times$reg & 2.91 & 0.24 & \textcolor{black}{0.44} & \textcolor{black}{2.97} & \textcolor{black}{0.27} & \textcolor{black}{0.48} & \textcolor{black}{0.70} & \textcolor{black}{0.49} & \textcolor{black}{0.32} & \textcolor{black}{0.87}\\
educ$\times$white & 10.80 & 0.28 & \textcolor{black}{0.51} & \textcolor{black}{10.64} & \textcolor{black}{0.21} & \textcolor{black}{0.25} & \textcolor{black}{0.09} & \textcolor{black}{0.42} & \textcolor{black}{1.00} & \textcolor{gray}{2.69}\\
midwest$\times$white$\times$educ & 1.54 & 0.44 & \textcolor{black}{0.52} & \textcolor{black}{0.96} & \textcolor{black}{0.38} & \textcolor{black}{0.47} & \textcolor{black}{0.36} & \textcolor{black}{0.27} & \textcolor{black}{0.35} & \textcolor{black}{0.56}\\
midwest$\times$edu$\times$race & 1.68 & 0.10 & \textcolor{black}{0.07} & \textcolor{black}{0.69} & \textcolor{black}{0.04} & \textcolor{black}{0.06} & \textcolor{black}{0.04} & \textcolor{black}{0.05} & \textcolor{black}{0.88} & \textcolor{black}{1.01}\\
\bottomrule
\end{tabular}
}
\end{center}
 {\emph{Note:} Absolute error in the distribution of categorical variables, weighted by the target population proportion for each level.  Gray numbers indicate the variable was included as a calibration constraint, and so imbalances very near zero are expected.  Note that all interactions with education use a three-way education coding.}
\end{table}

In the lower rows of Table~\ref{tab:abs_error}, we investigate the balance on important interactions, including an one that \citet{kennedy2018evaluation} deemed important: midwest $\times$ education level $\times$ race (bottom row).  Without explicitly incorporating knowledge about the importance of these variables, \kpop\ significantly improves balance on this interaction, reducing the mean absolute error from the initial value of 1.68 down to 0.1 percentage point - a much greater reduction than any of the non-\kpop\ estimators.  When incorporating the mean first requirements \textit{kpop\ + mf} also effectively addresses this interaction, reducing absolute bias to between 0.04 to 0.07 percentage points.  We see similar patterns of improvements in performance of the \kpop\ methods across a number of important interactions. Notably, in each case, the \kpop\ \textit{+ mf} method outperforms its mean calibration counterpart, emphasizing the ``no worse'' nature of the mean-first approach.  Additionally, balance is achieved regardless of the specified mean first constraints, highlighting the robustness of \kpop\ to standard user-specified constraints.

\paragraph{Weight severity.}

The additional constraints solved by \kpop\ weights can lead to reduced effective sample sizes compared to other approaches. To calculate effective sample size, we use the Kish formulation of $\frac{\left(\sum_i w_i\right)^2}{\sum_i w_i^2}$. Here, \textit{post-stratification} and \textit{mean calibration (all)} have effective sample sizes of 983 and 1101, respectively, while \kpop\ and \textit{kpop + mf (all)} have effective sample sizes of 789 and 749 respectively. Similarly, the (minimum) number of observations required to arrive at 90\% of the total weight is 1235 and 1362 for \textit{post-stratification} and \textit{mean calibration (all)} respectively, but 1193 and 1223 for \kpop\ and \textit{kpop + mf (all)}. Thus, a price is paid for the \kpop's ability to balance on more general functions of $\X$, but it is a fairly modest one here.

\paragraph{Weighting diagnostics.} Improvements in balance on $\K$ can be assessed using the diagnostics described above in section \ref{sec:kernel_diag}. The $L_1$ gap between the kernel-based estimates of multivariate density fell from 0.0318 prior to weighting to 0.0011 or below under all \kpop\ estimates, roughly a 29-fold improvement. Similarly the bias bound showed 5-6 fold improvements under each set of weights as compared to the unweighted bias bound.  Additional diagnostic and descriptive results can be found in Appendix~\ref{app:dist_w}.

\paragraph{Estimates.} Recall that our target is a two-way vote difference on reported vote choice (D-R) of 2.48 percentage points and that, for each weighting model, we evaluate the average of the predicted ``retrospective vote choice.'' The Pew survey, without weights, is extremely non-representative, with an unweighted average Clinton two-party vote margin of -0.194 [-3.24, 2.85] percentage points (95\% confidence interval in brackets). Mean calibration on basic demographics, excluding educational attainment, flips the signs of the estimate, producing an estimated margin to 6.49 [5.71, 7.28] percentage points. Mean calibration including education performs well, with an estimate of 3.04 [2.26, 3.82]. This is consistent with the findings of \citet{kennedy2018evaluation} that educational attainment was a significant driver of both nonresponse and voting for Donald Trump, especially among white voters. Moving to mean calibration on all auxiliary variables, the estimate moves farther from the truth to 4.31 [3.72, 4.91]. Finally, mean calibration on the retrospectively informed choice of variables and the potentially important interaction of region and education among white voters generated an improved point estimate of 3.30 [2.66, 3.93] substantially closer to the truth. 

The \kpop\ estimates are both stable and close to the truth across different specifications. Using \kpop\ alone results in a weighted estimate of 2.50 [1.96, 3.04] percentage points, the closest to the truth of all nine methods tested. $kpop+mf$ with additional mean first calibration produces point estimates of 2.95 [2.45, 3.44] $(demos)$, 3.09 [2.51, 3.67] \emph{(with education)}, and 3.19 [2.59, 3.79] percentage points for $(all)$, all close to the target margin in the CCES. Appendix Table \ref{app:applres_num} summarizes these results across all weighting methods.

We note that the standard errors (and thus confidence intervals) for \kpop\ are not necessarily larger than, and in fact are often smaller than, those of other methods. On the one hand, \kpop\ may lead to more variable weights than other approaches, which can contribute to larger variance estimates. Simultaneously, however, the linearization/residualization standard errors we employ (\citealp{fuller1975regression, kott2016}) from the outcome any signal that can be explained linearly by the $\phi(X)$ calibrated upon. Thus,  when a significant component of the outcome variance can be explained by $\phi(X)$, this leads to a substantial reduction in the estimated standard error. This reduction may be more than sufficient to make up for potentially higher variance weights, resulting in overall shorter intervals. As demonstrated in Appendix \ref{sims:SEs}, these estimated standard errors have nominal coverage under simulation and closely reproduce the empirical standard deviation of estimates under resampling.

\begin{figure}[!ht]
    \begin{center}
            \includegraphics[width = 1\textwidth]{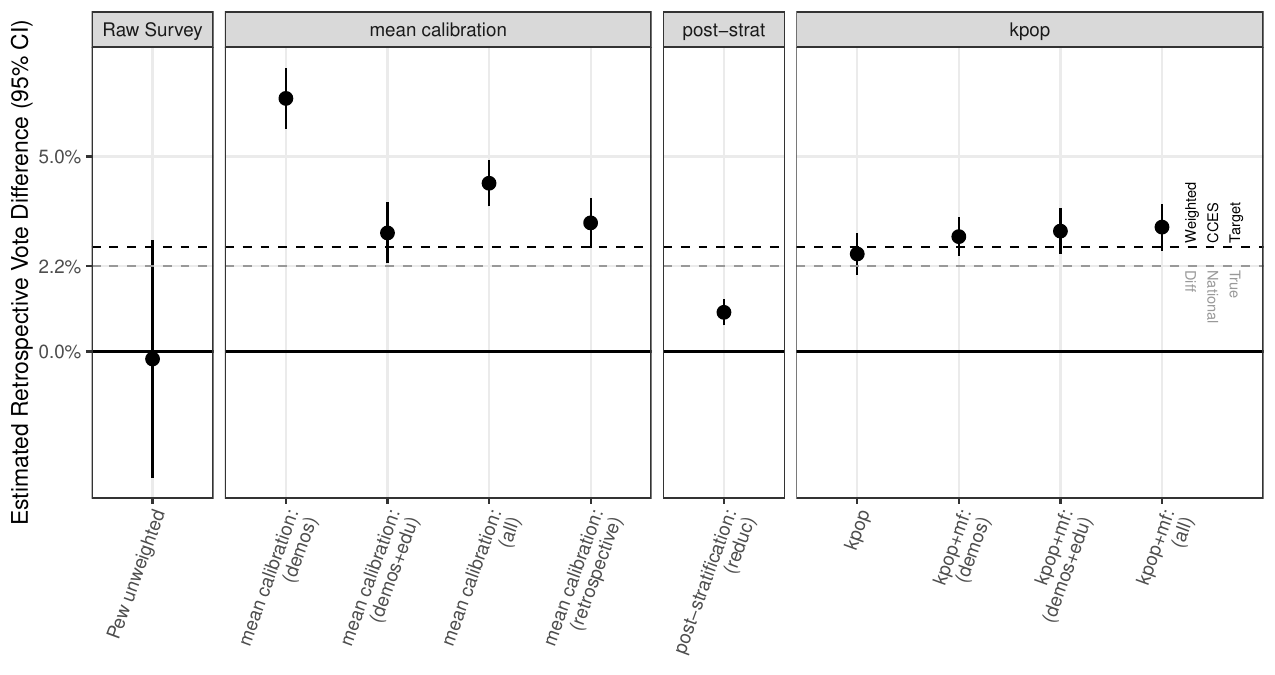} %
            \caption{{Comparison of weighting methods on Pew survey data to weighted CCES Target. Points represent the estimated vote margin from the predicted ``retrospective vote choice'' using Pew survey data and corresponding weighting scheme. The dashed black line indicates the target,  the reported two-way vote margin in the weighted CCES. The dashed grey line indicates the true values from national vote returns.}}
    \end{center}
\end{figure}

\section{Conclusions}\label{sec:discussion}
The challenges we seek to manage regarding common survey weighting techniques, particularly post-stratification and mean calibration, are well-known in the survey weighting literature (e.g. see \citealp{Kalton:2003vh}, \citealp{berinsky2006american}, \citealp{hartmanaccounting}).  Recent methods aim to address trade-offs between these approaches, as well as the relationship between mean calibration and inverse propensity score weighting \citep{linzer2011reliable, benmichael2021multilevel}. Variable selection for weighting has addressed one aspect of feature selection \citep{chen2019calibrating, mcconville2017model, caughey2017target}. Here we describe an approach that helps to reduce user discretion in the related problem of deciding what features and functions of observed covariates must be made to look similar in the sample and target population.

We note several limitations and areas for future work. First, the implementation described here makes no use of outcome data when constructing weights. This allows users to choose weights blind to the outcome to protect themselves against data ``snooping''. Further, these weights would be appropriate for estimating any outcome, as several are often of interest in a given survey. The downside, however, is that this leaves possible gains in efficiency and bias reduction on the table if there are functions of $X$ that predict response (and are thus imbalanced) but that do not predict the outcome and so need not be calibrated to achieve linear ignorability.  Still, recognizing such variables and choosing not to calibrate on them could lead to improved calibration on the remaining variables, possibly resulting in less extreme weights. Such an approach remains an option worth exploring in future work.

While we suggest the use of the Gaussian kernel with a variance maximizing choice of $b$, optimal selection of $b$ remains an area of ongoing research. Fortunately, our empirical results are not sensitive to the choice of $b$ (see Appendix~\ref{app:various_b}), but this may not always be the case. Further, while the present example focused on discrete $X$ for comparability to other approaches, a benefit of our approach, and the Gaussian kernel, is that it applies well---and perhaps more naturally---with continuous $X$. Nevertheless, other choices of kernels, and a means of choosing among them, is a fruitful area of future work.

Next, in our implementation, we use only the sample observations to form the columns of $\K$, a decision driven by feasibility constraints. Since the number of units in the target population is typically very large, using all sample and population units to form the columns of $\K$ is typically infeasible. Future work could consider ways of augmenting the columns of $\K$ by selecting population units that are poorly represented by sample units and using these additionally in the formation of the bases for calibration. Finally, our method relies on individual level population data, either from administrative data or a high-quality representative survey, which may not be accessible to all researchers.

In summary, \kpop\ is a kernel-based approach for weighting samples to be representative of target populations, while reducing reliance on user discretion and domain expertise to determine what covariates---and functions of those covariates---should be used for calibration. It does so by estimating a flexible, non-linear set of basis functions through a kernel transformation and achieving approximate balance on this representation of the covariates.  As shown in our application to the 2016 U.S. Presidential election, this method has great promise for reducing bias in non-representative samples.

\bibliography{main.bib}

@article{long1997regression,
  title={Regression models for categorical and limited dependent variables},
  author={Long, J.S.},
  journal={Advanced quantitative techniques in the social sciences},
  volume={7},
  year={1997},
  publisher={Sage Publications}
}

@book{hastie2009elements,
  title={The elements of statistical learning: data mining, inference, and prediction},
  author={Hastie, Trevor and Tibshirani, Robert and Friedman, Jerome H and Friedman, Jerome H},
  volume={2},
  year={2009},
  publisher={Springer}
}

@article{wong2018kernel,
  title={Kernel-based covariate functional balancing for observational studies},
  author={Wong, Raymond KW and Chan, Kwun Chuen Gary},
  journal={Biometrika},
  volume={105},
  number={1},
  pages={199--213},
  year={2018},
  publisher={Oxford University Press}
}

@article{yeying2018kernel,
  title={A Kernel-Based Metric for Balance Assessment},
  author={Yeying, Zhu and Debashis, Ghosh and others},
  journal={Journal of Causal Inference},
  volume={6},
  number={2},
  year={2018},
  publisher={De Gruyter}
}

@article{kallus2020generalized,
  title={Generalized optimal matching methods for causal inference.},
  author={Kallus, Nathan},
  journal={Journal of Machine Learning Research},
  volume={21},
  number={62},
  pages={1--54},
  year={2020}
}

@article{tarr2021estimating,
  title={Estimating average treatment effects with support vector machines},
  author={Tarr, Alexander and Imai, Kosuke},
  journal={arXiv preprint arXiv:2102.11926},
  year={2021}
}

@article{micchelli2006universal,
  title={Universal kernels},
  author={Micchelli, Charles A and Xu, Yuesheng and Zhang, Haizhang},
  journal={Journal of Machine Learning Research},
  volume={7},
  number={Dec},
  pages={2651--2667},
  year={2006}
}

@article{hainmueller2012entropy,
  title={Entropy balancing for causal effects: A multivariate reweighting method to produce balanced samples in observational studies},
  author={Hainmueller, Jens},
  journal={Political analysis},
  pages={25--46},
  year={2012},
  publisher={JSTOR}
}

@article{hazlett2020kernel,
  title={Kernel Balancing: A flexible non-parametric weighting procedure for estimating causal effects},
  author={Hazlett, Chad},
  journal={Statistica Sinica},
  year={2020},
  volume={30},
  pages={1155-1189}
}

@article{krls,
  title={Kernel regularized least squares: Reducing misspecification bias with a flexible and interpretable machine learning approach},
  author={Hainmueller, Jens and Hazlett, Chad},
  journal={Political Analysis},
  volume={22},
  number={2},
  pages={143--168},
  year={2014},
  publisher={Cambridge University Press}
}

@misc{kennedy_hartig_2019, title={``{R}esponse rates in telephone surveys have resumed their decline''},
    url={https://www.pewresearch.org/fact-tank/2019/02/27/response-rates-in-telephone-surveys-have-resumed-their-decline/},
    publisher={Pew Research Center}, 
    author={Kennedy, Courtney and Hartig, Hannah}, 
    year={2019}, 
    month={Feb}
}

@article{berinsky2006american,
  title={American public opinion in the 1930s and 1940s: The analysis of quota-controlled sample survey data},
  author={Berinsky, Adam J},
  journal={International Journal of Public Opinion Quarterly},
  volume={70},
  number={4},
  pages={499--529},
  year={2006},
  publisher={Oxford University Press England}
}

@book{sarndal2005estimation,
  title={Estimation in surveys with nonresponse},
  author={S{\"a}rndal, Carl-Erik and Lundstr{\"o}m, Sixten},
  year={2005},
  publisher={John Wiley \& Sons}
}

@article{Wu:2016ba,
    author = {Wu, Changbao and Lu, Wilson W},
    title = {{Calibration Weighting Methods for Complex Surveys}},
    journal = {International Statistical Review},
    year = {2016},
    volume = {84},
    number = {1},
    pages = {79--98},
    month = {April}
}

@article{Deville:1992bv,
    author = {Deville, Jean-Claude and S{\"a}rndal, Carl-Erik},
    title = {{Calibration Estimators in Survey Sampling}},
    journal = {Journal of the American Statistical Association},
    year = {1992},
    volume = {87},
    number = {418},
    pages = {376--382},
    month = jun
}

@article{Kalton:2003vh,
    author = {Kalton, Graham and Flores-Cervantes, Ismael},
    title = {{Weighting Methods}},
    journal = {Journal of Official Statistics},
    year = {2003},
    volume = {19},
    number = {2},
    pages = {81--97}
}

@article{Sarndal:2007vj,
    author = {S{\"a}rndal, Carl-Erik},
    title = {{The calibration approach in survey theory and practice}},
    journal = {Survey Methodology},
    year = {2007},
    volume = {33},
    number = {2},
    pages = {99--119},
    month = dec
}

@book{caughey_elements_2020,
  place={Cambridge},
  series={Elements in Quantitative and Computational Methods for the Social Sciences},
  title={Target Estimation and Adjustment Weighting for Survey Nonresponse and Sampling Bias},
  publisher={Cambridge University Press},
  author={Caughey, Devin and Berinskey, Adam J. and Chatfield, Sara and Hartman, Erin and Schickler, Eric and Sekhon, Jasjeet S.},
  year={2020},
  collection={Elements in Quantitative and Computational Methods for the Social Sciences}}

@article{Kott:2010dq, 
year = {2010}, 
title = {{Using Calibration Weighting to Adjust for Nonignorable Unit Nonresponse}}, 
author = {Kott, Phillip S and Chang, Ted}, 
journal = {Journal of the American Statistical Association}, 
doi = {10.2307/27920149?refreqid=search-gateway:2795e1b65b6b8b02c533c05b078cd6e2}
}

@techreport{pew2016,
     title = {{As Election Nears, Voters Divided Over Democracy and `Respect'}},
     year = {2016},
     author = {{Pew Research Center}},
     institution = {{Pew Research Center}},
     month = {10}
}

@article{Mercer:2017gb,
    author = {Mercer, Andrew W and Kreuter, Frauke and Keeter, Scott and Stuart, Elizabeth A},
    title = {{Theory and Practice in Nonprobability Surveys}},
    journal = {Public Opinion Quarterly},
    year = {2017},
    volume = {81},
    number = {S1},
    pages = {250--271},
    month = apr
}

@article{kennedy2018evaluation,
  title={An evaluation of the 2016 election polls in the United States},
  author={Kennedy, Courtney and Blumenthal, Mark and Clement, Scott and Clinton, Joshua D and Durand, Claire and Franklin, Charles and McGeeney, Kyley and Miringoff, Lee and Olson, Kristen and Rivers, Douglas and others},
  journal={Public Opinion Quarterly},
  volume={82},
  number={1},
  pages={1--33},
  year={2018},
  publisher={Oxford University Press US}
}

@misc{DVN/GDF6Z0_2017,
    author = {Ansolabehere, Stephen and Schaffner, Brian F.},
    publisher = {Harvard Dataverse},
    title = {{CCES Common Content, 2016}},
    UNF = {UNF:6:WhtR8dNtMzReHC295hA4cg==},
    year = {2017},
    version = {V4},
    doi = {10.7910/DVN/GDF6Z0},
    url = {https://doi.org/10.7910/DVN/GDF6Z0}
}

@book{little2019statistical,
  title={Statistical analysis with missing data},
  author={Little, Roderick JA and Rubin, Donald B},
  volume={793},
  year={2019},
  publisher={John Wiley \& Sons}
}

@article{zhao2016entropy,
  title={Entropy balancing is doubly robust},
  author={Zhao, Qingyuan and Percival, Daniel},
  journal={Journal of Causal Inference},
  volume={5},
  number={1},
  year={2016},
  publisher={De Gruyter}
}

@incollection{hartmanaccounting,
  author       = {Erin Hartman and Ines Levin},
  title        = {Accounting for Complex Survey Designs: Strategies for Post-stratification and Weighting of Internet Surveys},
  booktitle    = {Oxford Handbook of Electoral Persuasion},
  publisher    = {Oxford University Press},
  year         = 2019,
  editor       = {Elizabeth Suhay and Bernard Grofman and Alexander H. Trechsel},
  pages        = {},
  address      = {},
  edition      = {},
  month        = {}
}

@misc{benmichael2021multilevel,
      title={Multilevel calibration weighting for survey data}, 
      author={Eli Ben-Michael and Avi Feller and Erin Hartman},
      year={2021},
      eprint={2102.09052},
      archivePrefix={arXiv},
      primaryClass={stat.ME}
}

@article{linzer2011reliable,
  title={Reliable inference in highly stratified contingency tables: Using latent class models as density estimators},
  author={Linzer, Drew A},
  journal={Political Analysis},
  pages={173--187},
  year={2011},
  publisher={JSTOR}
}

@article{chen2019calibrating,
  title={Calibrating non-probability surveys to estimated control totals using LASSO, with an application to political polling},
  author={Chen, Jack Kuang Tsung and Valliant, Richard L and Elliott, Michael R},
  journal={Journal of the Royal Statistical Society: Series C (Applied Statistics)},
  volume={68},
  number={3},
  pages={657--681},
  year={2019},
  publisher={Wiley Online Library}
}

@article{mcconville2017model,
  title={Model-assisted survey regression estimation with the lasso},
  author={McConville, Kelly S and Breidt, F Jay and Lee, Thomas CM and Moisen, Gretchen G},
  journal={Journal of Survey Statistics and Methodology},
  volume={5},
  number={2},
  pages={131--158},
  year={2017},
  publisher={Oxford University Press}
}

@article{caughey2017target,
  title={Target Selection as Variable Selection: Using the Lasso to Select Auxiliary Vectors for the Construction of Survey Weights},
  author={Caughey, Devin and Hartman, Erin},
  journal={Available at SSRN 3494436},
  year={2017}
}

@article{kott2016,
author = {Kott, Phillip S.},
title = {Calibration weighting in survey sampling},
journal = {WIREs Computational Statistics},
volume = {8},
number = {1},
pages = {39-53},
doi = {https://doi.org/10.1002/wics.1374},
url = {https://wires.onlinelibrary.wiley.com/doi/abs/10.1002/wics.1374},
eprint = {https://wires.onlinelibrary.wiley.com/doi/pdf/10.1002/wics.1374},
year = {2016}
}

@article{fuller1975regression,
  title={Regression analysis for sample survey},
  author={Fuller, Wayne A},
  journal={Sankhya},
  volume={37},
  number={3},
  pages={117--132},
  year={1975}
}

@article{hartman_huang_2023, 
    title={Sensitivity Analysis for Survey Weights}, 
    DOI={10.1017/pan.2023.12}, 
    journal={Political Analysis}, 
    publisher={Cambridge University Press}, 
    author={Hartman, Erin and Huang, Melody}, 
    year={2023}, 
    pages={1–16}}

@article{opsomer2021replication,
  title={Replication variance estimation after sample-based calibration.},
  author={Opsomer, Jean D and Erciulescu, Andreea L},
  journal={Survey Methodology},
  volume={47},
  number={2},
  pages={265--278},
  year={2021},
  publisher={Statistics Canada}
}


\clearpage
\appendix
\setcounter{page}{1}
\renewcommand\thefigure{\thesection.\arabic{figure}}    
\setcounter{figure}{0}
\renewcommand\thetable{\thesection.\arabic{table}}    
\setcounter{table}{0}

\begin{center}
    {\Large Supplementary Materials for ``\kpop: A kernel balancing approach for reducing specification assumptions in survey weighting''}
\end{center}
\appendix
\singlespacing

\section{Proofs}\label{app:proofs}

To first review notation, we consider covariates $X_i$, the outcome of interest, $Y_i$, and an indicator for having completed data on $Y_i$, $R_i \in \{0,1\}$, accounting for both unit $i$ being in a sample and completing the question.  In addition, we consider an \emph{unobserved} factor $Z_i$.  Independent observations of the form $\{X_i, Z_i, Y_i, R_i \}$ are realized from a common joint density $p(X_i,Z_i,Y_i,R_i)$. Investigators also have a smaller survey respondent sample, consisting only of $\Ns$ units. We consider first the case where these $\Ns$ sampled units are drawn, with bias, from the $\Np$ units in the target population data, with $R_i=1$ indicating sampled units, who thus have non-missing values of $Y_i$. Below we also consider the case when the sample is drawn disjointly from the target population.

\subsection{Outcome linearity} \label{app:proof_linear_outcome}

While the full linear ignorability assumption defined in the paper regards the independence of (link) linear residuals from both the outcome and response models, to develop the underlying ideas we will first consider only the residual in the outcome model and how it relates to the full response process, not just the residual in the response process.

The decomposition of $Y$ into $\phi(X)^\top\beta + \epsilon$ is always possible, and so the assumptions embodied by this decomposition are borne out by further assumptions on $\epsilon$.   Here, unlike in many regression settings, the concern is not the relationship between $\phi(X)$ and an $\epsilon$ defined as other causes of $Y$, but rather between the $\epsilon$ that is constructed simply to be orthogonal to $X$, and the sample-presence indicator, $R$. 

While non-parametric ignorability asserts that $Y \indep R \mid X$,  suppose we replace this with the notion that the variation in $Y$ not explained linearly by $\phi(X)$, namely $\epsilon$, is independent of $R$: $\epsilon \indep R$. Note that (i) the non-parametric conditioning on $\X$ has been replaced here by a specific parametric adjustment, $Y-\phi(X)^\top\beta$, and (ii) we have not yet removed the linear relationship between $\phi(X)$ and $R$, and doing so is complicated by $R$'s potential link-linear relationship to $\phi(X)$, but will consider that shortly. 

Without loss of generality,  consider the decomposition of $Y$ as
\[ Y = \phi(X)^\top\beta + Z + \nu \]

\noindent in which $Z$ can take various forms that will affect whether linear ignorability holds and such that $\nu$ contains merely idiosyncratic noise independent of $\phi(X)$, $Z$, and $R$. The $\epsilon$ above is now composed of $Z+\nu$. Note we can assume $Z$ is orthogonal to $\phi(X)$ because, if it wasn't, one could replace the original $Z$ with the component not in the span of $\phi(X)$, $Z-(\phi(X)^{\top}\phi(X))^{-1}\phi(X)^{\top}Z$. This will change the $\beta$ of the best fitting model, but that is not a concern here. 

The weighted average outcome in the sample will then be,

\begin{align*}
\mu_s(w) = \frac{1}{\Ns}\sum Y_i =  \frac{1}{\Ns} \sum w_i \phi(\X)_i^{\top}\beta + \frac{1}{\Ns} \sum w_i Z_i + \frac{1}{\Ns} \sum w_i \nu_i 
\end{align*}

\noindent while the mean outcome in the target population is 

\begin{align*}
\mu = \overline{\phi(X)}_{pop}^{\top}\beta + \frac{1}{N_{pop}} \sum Z_i + \nu_i
\end{align*}

\noindent which will not equal $\E[Y]$ exactly, but $p(\cdot)$ is constructed such that $\E[\mu]=\E[Y]$.  The bias under exact mean calibration on $X$ is then

\begin{align*}
    \text{bias} &= \E[\mu_s(w) - \mu] \\
    &= \E \left[ \frac{1}{\Ns} \sum w \phi(X_i)^{\top}\beta + \frac{1}{\Ns} \sum w_i Z_i + \frac{1}{\Ns} \sum w_i \nu_i  - \left(\overline{\phi(X)}_{pop}^{\top}\beta + \frac{1}{N_{pop}} \sum Z_i + \frac{1}{N_{pop}} \sum \nu_i \right) \right] \\
    &= \E \left[ \overline{\phi(X)}_{pop}^{\top}\beta + \frac{1}{\Ns} \sum w_i Z_i + \frac{1}{\Ns} \sum w_i \nu_i  - \left(\overline{\phi(X)}_{pop}^{\top}\beta + \frac{1}{N_{pop}} \sum Z + \frac{1}{N_{pop}} \sum \nu_i \right) \right] \\
&= \E \left[ \frac{1}{\Ns} \sum w_i Z_i  - \left( \frac{1}{N_{pop}} \sum Z_i \right) \right]  + \E\left[\frac{1}{\Ns} \sum w_i \nu_i - \frac{1}{N_{pop}} \sum \nu_i \right]\\
&= \E[wZ\mid R=1] - \E[Z] + 0 \\
\end{align*}

Hence if $Z$ (and thus $\epsilon=Z+\nu$) is not orthogonal to $R$, the bias is simply \textit{the difference between the weighted average of $Z$ in the sample and the average of $Z$ in the target population.} Note that while $\E[\cdot]$ here integrates over $p$ (the full data-generating process), this is an expectation of the bias as defined by the difference between a fixed target population mean, i.e. the $\mu$ we would have versus our estimate of it from a smaller weighted sample, $\mu_s(w)$.

So far this result does not depend on whether the sample units ($i:R_i=1$) are a subset of the target population units. Note that the term $\left[\frac{1}{\Ns} \sum_{i:R=1} w_i \nu_i - \frac{1}{N_{pop}} \sum_{i} \nu_i \right]$ equals zero in expectation so disappears from the bias but is one source of error in a finite sample. In the case where the sample is a subset of the target population, the second term in this expression can be rewritten as a combination of the sample units and the units in the sample not in the population, leaving us with 

\begin{align}
 &= \frac{1}{\Ns} \sum_{i:R=1} w_i \nu_i - \left(\frac{\Ns}{\Np}\right) \frac{1}{\Ns} \sum_{i:R=1} w_i \nu_i + \left(\frac{\Np-\Ns}{\Np}\right) \frac{1}{\Np-\Ns} \sum_{i:R_i=0} \nu_i \\
 &= \frac{1}{\Ns} \sum_{i:R=1} w_i \nu_i - \frac{1}{\Ns} \sum_{i:R=1} w_i \nu_i +\frac{1}{\Np} \sum_{i:R=0} \nu_i \\
 &= \frac{1}{\Np} \sum_{i:R=0} \nu_i
\end{align} 

\noindent Hence, this particular source of finite sample error is smaller when the sample is a subset of the target population: because the sample and population share units, they differ on $\nu$ only for the observations that are in the target population and not the sample.

When we consider a $Z \indep R$ then $\epsilon = Z+\nu$ is independent of $R$. Accordingly, when $Z \indep R$, $\E[Z|R=1]=\E[Z]$. Further, as the weights depend upon only $R$ and $X$, $Z \indep X$ and $Z \indep R$ implies $Z \indep w$. Hence, when Assumption~\ref{assumption.linearign} holds, the final line above becomes $\E[w | R=1]\E[Z | R=1]-\E[Z] = \E[Z]-\E[Z]=0$, concluding our proof.

\subsection{Link-linearity in the sampling process}\label{app:proof_linear_sampling}

The full linear ignorability assumption (\ref{assumption.linearign}) is met not only when $\epsilon \indep R$, but more weakly when $\epsilon \indep \eta$, where $\eta$ is the part of the sampling probability that cannot be explained (link) linearly by $\phi(X)$. Derivations working with $\eta$ are complicated by the fact that, as a binary variable, we do not expect to model $R$ directly as a linear function of $\phi(X)$ but rather through a link function. Suppose we impose a general link-linear decomposition, $Pr(R=1|X)= g(\phi(X)^{\top}\theta + \eta)$, where $g(\cdot)$ is some (inverse) link function $g: \mathbb{R} \mapsto [0,1]$. 
We now consider the assumption of independence only between the linear residual of the outcome and ``link-linear'' residual for the selection, are independent, i.e. that $\epsilon \indep \eta$. 

In the ``nested'' case used in the text, where the sample contains a subset of the units in the target population, mean calibration weights solve the moment conditions,
\begin{align}\label{eq:phiconstraints}
    \sum_{x \in X} \phi(x) w(x) p(X=x \mid R=1) = \sum_{x \in X} \phi(x) p(X=x) 
\end{align}
\noindent where $p(X=x)$ is the distribution among the target population. Note that though $\phi(X)$ has an infinite-dimensional representation in the case of the Gaussian kernel, the virtue of the kernel representation is that this can be achieved by calibrating on $K_i$ instead of $\phi(X_i)$.  The number of constraints to be solved (neglecting the  approximation) is thus the number of dimensions of $K_i$, namely $\Ns$. Further, the same moment conditions would be solved were we to choose weights according to the function $w(x) = \frac{p(X=x)}{p(X=x|R=1)}$. We thus assume the empirical weights solving Equation \ref{eq:phiconstraints} converge to those described by $w(x)$ and can thus be expressed as
\begin{align}\label{weights.nested}
    w(x) &= \frac{p(x)}{p(x \mid R=1)} =  \frac{p(x)p(R=1)}{p(R=1|x)p(x)} =\frac{p(R=1)}{p(R=1 |x)}
\end{align}

\noindent producing the well-known ``inverse probability of selection'' weights. In particular the sample weights become 

\begin{align}
    w_i &= \frac{\Ns/\Np}{ g(\phi(X_i)^{\top}\theta + \eta_i)} 
\end{align}

The ``non-nested' case, where the sample ($R=1$) is drawn not as a subset of the target population but as a separate group, is similar. Here, mean calibration for $\phi(x)$ solves
\begin{align*}
    \sum_{x \in X} \phi(x) w(x) p(X=x \mid R=1) = \sum_{x \in X} \phi(x) p(X=x \mid R=0)   \\
\end{align*}
\noindent where $R=0$ designates units in the target population alone and $R=1$ are those in the sample alone. Expression~(\ref{weights.nested}) is then replaced by
\begin{align}\label{weights.nonnested}
    w(x) &= \frac{p(x \mid R=0)}{p(x \mid R=1)} = \frac{p(R=0|x)p(R=1)}{p(R=1 |x) p(R=0)} = \frac{p(R=0|x)}{p(R=1 |x)}\frac{\Ns}{\Np} 
\end{align}

\noindent so that $w(X_i)$ is just a rescaling of $\frac{1-g(\phi(X_i)^{\top}\theta + \eta_i)}{g(\phi(X_i)^{\top}\theta + \eta_i)}$, i.e. the inverse ``odds of selection'' rather than probability of selection. In either case, the weights are a function of  $X_i^{\top}\theta + \eta_i$, which we rename $h(X_i^{\top}\theta + \eta_i)$. For example, supposing $g(\cdot)$ was the inverse link function for the logit, then in the non-nested case,

\[ w_i = h(\phi(X_i)^{\top}\theta + \eta_i) = \frac{1-g(\phi(X_i)^{\top}\theta + \eta_i)}{ g(\phi(X_i)^{\top}\theta + \eta_i)} \frac{\Ns}{\Np}= e^{-(\phi(X_i)^{\top}\theta + \eta_i)} \frac{\Ns}{\Np} \]

However the logit link is just one example; we can proceed for any choice of $g(\cdot): \mathbb{R} \mapsto [0,1]$. As derived above, the bias from using any weights $w$ to approximate the mean outcome in the target population is given by $\E \left[\sum_i w_i Z_i \right] - \E[Z]$. Under linear ignorability, this is
\begin{align}
    \text{bias} &= \E \left[\sum_i w_i Z_i \right] - \E[Z] \\
     &= \E \left[\sum_i h(\phi(X_i)^{\top}\theta + \eta_i) Z_i \right] -  \E[Z]\\
    &= \E[Z] \Ns \E \left[h(\phi(X_i)^{\top}\theta+\eta_i) \right] - \E[Z] \label{blah} \\
    &= \E[Z] \Ns \E[w_i] - \E[Z] \\
    &= 0
\end{align}

\noindent where (\ref{blah}) follows from the prior line because (i) $Z$ is orthogonal to $\phi(X)$ and thus any function $\phi(X)^{\top}\theta$, and (ii) linear ignorability ensures that $Z \indep \eta$.

\clearpage

\clearpage

\section{Detailed choices of kernel and data scaling}\label{app:kernelchoice}

We employ the Gaussian kernel throughout and in our software implementation. Users could reasonably choose other kernels that form function spaces more suitable to their purposes. The Gaussian kernel, however, is a useful default choice for several reasons. The choice of $\phi(\cdot)$ corresponding to the Gaussian kernel is infinite dimensional, and the Gaussian is one of several ``universal'' kernels, able to fit any continuous function given enough observations (see e.g. \citealp{micchelli2006universal}). More importantly, in a given finite sample scenario, the functions that are fitted by Gaussian kernels---those that can be built by rescaling and adding Gaussian curves over all the observations in the sample---form an appealing range of smooth, flexible functions suitable for many applications.

In estimating survey weights using \kpop, several data pre-processing decisions must be made. We provide default recommendations to reduce the number of user-driven decisions. For continuous variables, we recommend rescaling so that all variables have variance one. This avoids unit-of-measure decisions from affecting the relative influence of different variables in the similarity measure between units. In our application, however, for comparability to other survey weighting approaches, we focus on categorical variables. Gaussian kernels remain appropriate with categorical variables, as demonstrated in the  minimal example in \ref{sec:minimal_example}, but require different data pre-processing decisions. Differences between units in a categorical setting are binary: two units are either an exact match or not. We therefore use dummy or ``one-hot encoding'' of categorical variables \textit{without dropping any levels} and do not rescale the resulting binary variables by their standard deviations. Under these choices, the numerator of the Gaussian kernel $||X_i - X_j||^2$ is simply two times the number of variables on which units $X_i$ and $X_j$ differ, regardless of the number of levels attached to each of those variables. 

This brings us to the choice of $b$ in the kernel definition. This is effectively a feature extraction choice, constrained by feasibility. Obtaining balance on a kernel matrix formed with a small $b$ implies finer balance and less interpolation, but at the cost of more extreme weights (and eventually, infeasibility). To see how $b$ affects the information in the kernel matrix in more detail, see the following section which displays various scree plots by the choice of $b$.

We propose a new approach to selecting $b$. In related kernel methods, a common default choice for $b$ is often $b=D$ or $b=2D$ where $D$ is the number of dimensions of $\X$ used to form $\K$. One reason is that, for standardized data, the expectation of the numerator in the exponent grows proportionally to $D$ \citep{krls}. Hence scaling proportionally to $D$ is reasonable, keeping the magnitude of the exponent stable as $D$ grows. However the best constant of proportionality remains unclear. Recall that $b$ scales the similarly measure $||X_i - X_j||^2$. There is no clear optimal value of $b$, without further assumption. Instead, we seek a reasonable choice that is computationally convenient and avoids researcher intervention. A choice of $b$ is reasonable, we argue, if it produces a range of values in each column of $\K$, encoding meaningful variability in the similarity/difference between different pairs of units. Given that this measure is bounded between 0 and 1, we use the variance of $\K$ (disregarding the diagonal, which is unchanging) as a measure of the useful information available at a given choice of $b$. We then choose the variance-maximizing value of $b$. We make no claim as to the optimality of this result, but it does guarantee a reasonable choice is made, without user intervention and without inspecting $Y$. In our simulations and applications, this choice produces consistently good performance, though the results are shown to be stable across a wide range of $b$ regardless (see appendix \ref{app:various_b}).

\clearpage
\subsection{Singular value decomposition of kernel matrix}\label{app:scree_k}

The following plot displays the first 50 singular vectors and their corresponding scaled singular values. Singular values are scaled by the largest singular value to facilitate  comparisons across different $K$ matrices. As can be seen in the plot, smaller choice of $b$, the denominator in the Gaussian exponent, correspond to a larger number of meaningful singular vectors while larger values lead to fewer. This helps to illuminate how the choice of $b$ represents a feature extraction choice. Larger $b$ values smooth over smaller differences between units, resulting in a kernel matrix which contains fewer dimensions of meaningful variance. This can be seen in the sharp drop of in the magnitude of singular values by even the 10th singular vector. Smaller choice of $b$ on the other hand, maintained more fine-grained information regarding the difference between units. Accordingly, we see a less precipitous decline in meaningful singular values the smaller the choice of $b$, with even the 50th singular vector still corresponding to a non-zero singular value. In this way, the choice of $b$ represents a feature extraction choice. As $b$ grows increasingly small, this choice also becomes constrained by feasibility challenges as $\kpop$ must seek balance on a larger and larger number of meaningful singular vectors of $K$.

\begin{figure}[!htp]
    \centering
    \includegraphics[width=1\textwidth]{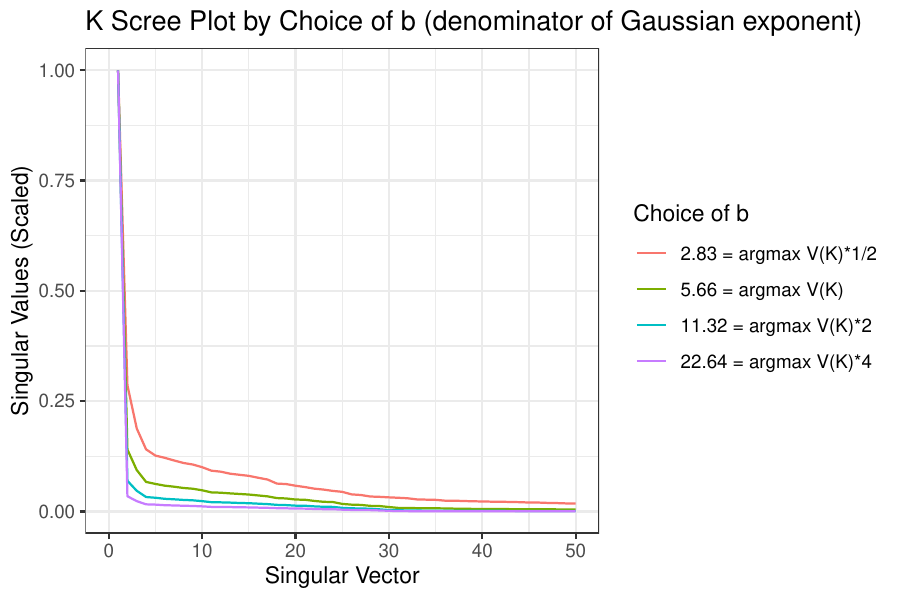}
    \caption{Screeplot of first 50 singular values of kernel matrix for different choices of $b$ (scaled by maximum singular value for comparability across matrices).  }
\end{figure}

\clearpage
\section{Application: Additional Details}

\subsection{Description of target population}
\renewcommand{\arraystretch}{1}
\begin{table}[ht!]
\caption{\label{tab:margins_table}Marginal distribution, in percentage points, of important demographics for the target population and the unweighted sample}
\centering
\begin{threeparttable}
\small
\begin{tabular}[t]{>{\raggedright\arraybackslash}p{1.45in}>{\raggedleft\arraybackslash}p{0.45in}>{\raggedleft\arraybackslash}p{0.45in}}
\toprule
& Target & unweighted\\
 & (CCES) & Pew\\
\midrule
\addlinespace[.5em]
\multicolumn{3}{l}{\textbf{Party Identification}}\\
\hspace{1em}Dem & 38.1 & 34.5\\
\hspace{1em}Ind & 32.3 & 35.0\\
\hspace{1em}Rep & 29.5 & 30.5\\
\multicolumn{3}{l}{\textbf{4-way Age Bucket}}\\
\hspace{1em}18 to 35 & 28.7 & 19.1\\
\hspace{1em}36 to 50 & 21.3 & 21.4\\
\hspace{1em}51 to 64 & 29.9 & 31.0\\
\hspace{1em}65+ & 20.1 & 28.5\\
\multicolumn{3}{l}{\textbf{Gender}}\\
\hspace{1em}Female & 50.9 & 47.2\\
\hspace{1em}Male & 49.1 & 52.8\\
\multicolumn{3}{l}{\textbf{Race/Ethnicity}}\\
\hspace{1em}Black & 11.7 & 8.8\\
\hspace{1em}Hispanic & 6.5 & 7.6\\
\hspace{1em}Other & 6.8 & 7.1\\
\hspace{1em}White & 75.0 & 76.5\\
\multicolumn{3}{l}{\textbf{Region}}\\
\hspace{1em}Midwest & 23.4 & 22.1\\
\hspace{1em}Northeast & 19.7 & 18.3\\
\hspace{1em}South & 35.4 & 37.9\\
\hspace{1em}West & 21.4 & 21.7\\
\multicolumn{3}{l}{\textbf{Education Level}}\\
\hspace{1em}No HS & 6.8 & 2.6\\
\hspace{1em}High school & 30.6 & 19.7\\
\hspace{1em}Some college & 23.0 & 15.2\\
\hspace{1em}2-year & 10.6 & 11.4\\
\hspace{1em}4-year & 18.7 & 29.1\\
\hspace{1em}Post-grad & 10.4 & 22.1\\
\multicolumn{3}{l}{\textbf{Income}}\\
\hspace{1em}Less than 20k & 10.5 & 9.2\\
\hspace{1em}20-50k & 29.9 & 25.3\\
\hspace{1em}50-100k & 32.5 & 27.2\\
\hspace{1em}100-150k & 11.5 & 15.2\\
\hspace{1em}More than 150k & 5.5 & 14.8\\
\hspace{1em}Prefer not to say & 10.1 & 8.3\\
\bottomrule
\end{tabular}
\end{threeparttable}
\end{table}

\renewcommand{\arraystretch}{1}
\begin{table}[ht!]
\ContinuedFloat
\caption{ Marginal distribution in percentage points (Cont.)}
\centering
\begin{threeparttable}
\small
\begin{tabular}[t]{>{\raggedright\arraybackslash}p{1.45in}>{\raggedleft\arraybackslash}p{0.45in}>{\raggedleft\arraybackslash}p{0.45in}}
\toprule
& Target & Unweighted\\
 & (CCES) & Pew\\
\midrule
\addlinespace[.5em]

\multicolumn{3}{l}{\textbf{Religious Affiliation}}\\
\hspace{1em}Catholic & 21.3 & 21.6\\
\hspace{1em}Jewish & 2.3 & 3.5\\
\hspace{1em}Not religious & 32.4 & 22.2\\
\hspace{1em}Other & 3.4 & 4.9\\
\hspace{1em}Protestant & 40.5 & 47.8\\
\multicolumn{3}{l}{\textbf{born-again Christian}}\\
\hspace{1em}No & 67.4 & 69.1\\
\hspace{1em}Yes & 32.6 & 30.9\\
\multicolumn{3}{l}{\textbf{Church Attendance}}\\
\hspace{1em}Never & 49.6 & 30.3\\
\hspace{1em}Weekly & 27.9 & 34.9\\
\hspace{1em}Monthly & 8.2 & 15.7\\
\hspace{1em}Yearly & 14.3 & 19.2\\
\bottomrule
\end{tabular}
\end{threeparttable}
\end{table}

\clearpage

\subsection{Details of predictive outcome model}\label{app:outcome_model} 

We run a regularized multinomial logistic model run on the CCES data to predict (weighted) three-way vote choice and include all variables in Table \ref{tab:margins_table} and a handful of interactions among them. Specifically, the model includes gender, 3-way party id, race/ethnicity, 6-way education, region, 6-way income, 5-way religion, 4-way church attendance, born-again status, continuous age, age$^2$, gender $\times$ party id, and finally age (cont.) $\times$ party id. We train this model on an 80\% sample of the CCES target population. We use 10-fold cross-validation to select $\lambda$, the regularization parameter, choosing the value  which produces cross-validation error that is within one standard error of the the minimum. When we classify vote choice as the highest predicted probability outcome for an individual, this model accurately classifies 79\% of vote choice in the test set with the following confusion matrix (in percent):
\begin{table}[h!]
\centering
\begin{tabular}{lrrr}
\toprule
  & & Reference & \\
  & Democrat & Other & Republican\\
Predicted & & & \\  
\midrule
\hspace{1em}Democrat & 86.530 & 47.098 & 14.049\\
\hspace{1em}Other & 0.067 & 0.792 & 0.000\\
\hspace{1em}Republican & 13.403 & 52.111 & 85.951\\
\bottomrule
\end{tabular}
\caption{Lasso multinomial model confusion matrix (in percent) for test set observations when classifying their predicted outcome among Democrat, Republican and Other by the highest modeled probability.}
\end{table}

When we classify test observations by their outcome with highest modeled probability, we have the most trouble accurately predicting those who choose ``Other'' which make up only 9\% of responses. This is of minimal concern, however, because we focus on outcomes which consider only the vote share between ``Democrat'' and ``Republican.'' In particular, we focus on the weighted average individual vote margin defined simply as the difference between the model's predicted probability of voting Democrat and Republican, $p(D) - p(R)$. We project this model onto the Pew sample data to predict three-way retrospective vote choice and subsequently compare the performance of different methods in weighting the sample's projected outcomes to our target. Recall that our target is the reported retrospective vote choice reported by respondents to the CCES.

\clearpage

\subsection{Numerical Performance on Modeled Retrospective Vote Difference in Application} \label{app:applres_num}

\begin{table}[h!]

\caption{Performance on Modeled p(D) - p(R) Vote Difference among Various Weighting Methods}
\begin{center}
\begin{tabular}[t]{lrrrr}
\toprule
  & Estimate & SE & CI Lower & CI Upper\\
\midrule
\textbf{CCES Target} & 2.68 & 0.51 & 1.68 & 3.67\\
\hspace{1em}Pew unweighted & -0.19 & 1.55 & -3.24 & 2.85\\
\hspace{1em}mean calibration (demos) & 6.49 & 0.40 & 5.71 & 7.28\\
\hspace{1em}mean calibration (d+edu) & 3.04 & 0.40 & 2.26 & 3.82\\
\hspace{1em}mean calibration (all) & 4.31 & 0.30 & 3.72 & 4.91\\
\hspace{1em}mean calibration retrospective & 3.30 & 0.32 & 2.66 & 3.93\\
\hspace{1em}post-stratification (reduced) & 1.00 & 0.17 & 0.66 & 1.34\\
\hspace{1em}post-strat (all) & -6.81 & 0.02 & -6.84 & -6.78\\
\hspace{1em}kpop & 2.50 & 0.27 & 1.96 & 3.04\\
\hspace{1em}kpop+mf (demos) & 2.95 & 0.25 & 2.45 & 3.44\\
\hspace{1em}kpop+mf (d+edu) & 3.09 & 0.30 & 2.51 & 3.67\\
\hspace{1em}kpop+mf (all) & 3.19 & 0.31 & 2.59 & 3.79\\
\bottomrule
\end{tabular}
\end{center}
{\emph{Note:} Standard Error estimators differ across estimates: for the CCES Target and the unweighted Pew estimate, the Horvitz-Thompson estimator is used; for all other estimates, the linearization SE estimator is used. }
\end{table}

\clearpage

\subsection{Distribution of Weights and Diagnostic Measures}
\label{app:dist_w}

Note that for ease of interpretability, all of the following use weights that sum to $\Ns$. In this application that corresponds to 2052 pew units.

\begin{figure}[!htp]
    \centering
    \includegraphics[width=\textwidth]{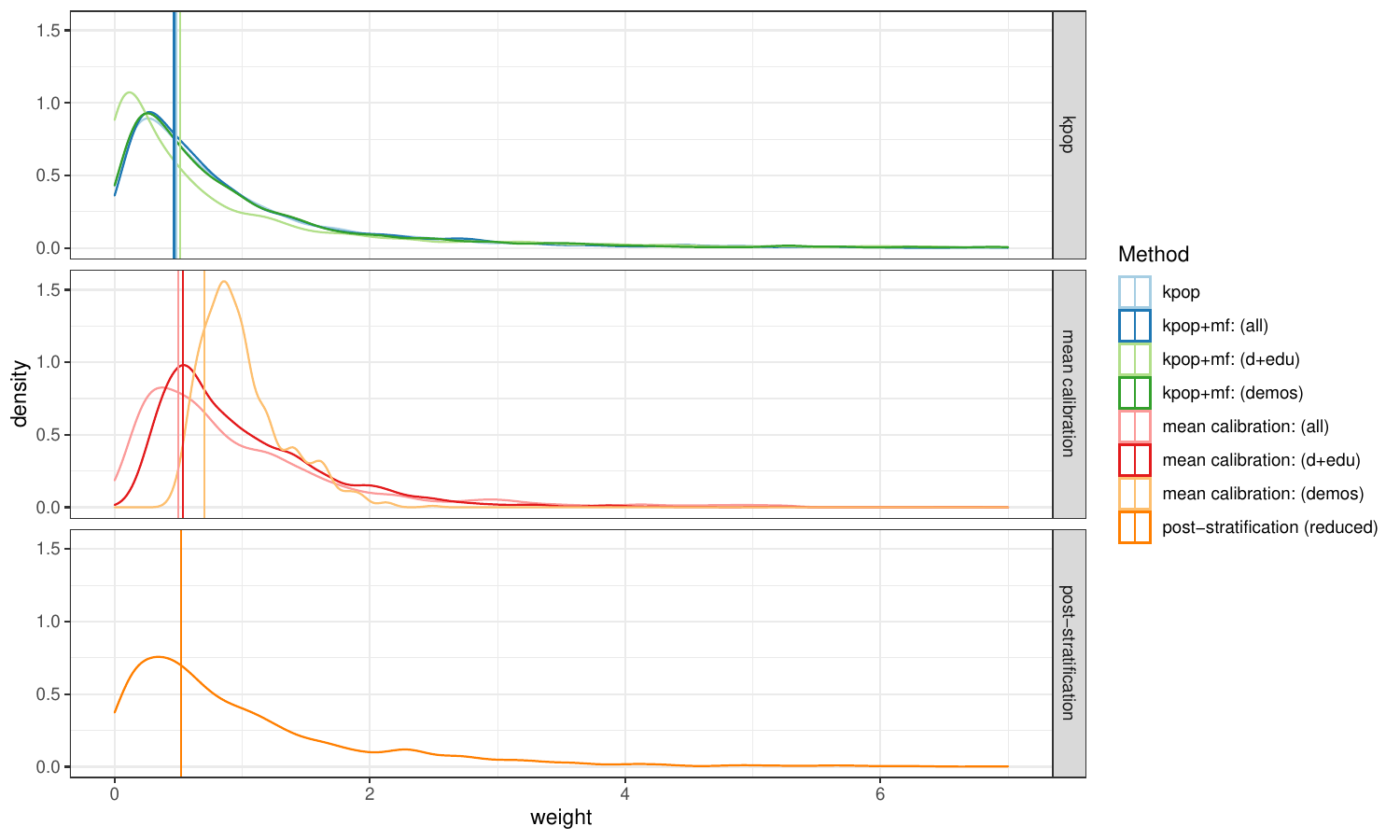}
    \caption{Distribution of weights (for Pew sample) across the various weighting methods discussed in Section~\ref{sec:application}, focusing on a smaller range of weight magnitudes (horizontal axis). Vertical lines indicate where 90\% of the total sum of the weights ordered from largest to smallest occurs. Thus, the area to the right of these lines accounts for 90\% of the mass of the weights.}
\end{figure}

\begin{table}
\caption{Numerical Summary of Distribution of Survey Weights}
\centering
\begin{tabular}[t]{lrrrrrr}
\toprule
    & Variance & Max & Min & IQR & Effective & No. Units 
  to\\
   &  &  &  &  &  SS & 90\% Sum of\\
   &  &  &  &  &  (Kish) & Total (2052)\\
\midrule
kpop & 1.60 & 15.23 & 0.0103 & 0.92 & 789 & 1193\\
kpop+mf: (demos) & 1.98 & 28.82 & 0.0005 & 0.93 & 689 & 1183\\
kpop+mf: (d+edu) & 2.99 & 24.22 & 0.0001 & 1.01 & 515 & 901\\
kpop+mf: (all) & 1.74 & 23.29 & 0.0230 & 0.91 & 749 & 1223\\
mean calibration: (demos) & 0.11 & 2.49 & 0.4367 & 0.42 & 1852 & 1721\\
mean calibration: (d+edu) & 0.44 & 5.28 & 0.1515 & 0.78 & 1425 & 1535\\
mean calibration: (all) & 0.86 & 7.32 & 0.0516 & 0.88 & 1101 & 1362\\
post-stratification: (reduced) & 1.09 & 9.52 & 0.0050 & 0.96 & 983 & 1235\\
\bottomrule
\end{tabular}
\end{table}

\begin{table}

\caption{L1 and Biasbound Results for all kpop Methods in 2016 Election Application}
\begin{center}
\begin{tabular}[t]{lrrrrr}
\toprule
  & L1  & L1  & Biasbound  & Biasbound  & Biasbound \\
  & Original & Optimal & Original  & Optimal &  Ratio\\
\midrule
kpop & 0.0318 & 0.0011 & 0.0446 & 0.0069 & 6.4236\\
kpop+mf: (demos) & 0.0318 & 0.0006 & 0.0446 & 0.0079 & 5.6577\\
kpop+mf: (d+edu) & 0.0318 & 0.0005 & 0.0446 & 0.0078 & 5.7173\\
kpop+mf: (all) & 0.0318 & 0.0010 & 0.0446 & 0.0085 & 5.2345\\
\bottomrule
\end{tabular}
\end{center}
\emph{Note:} Original refers to the L1 and biasbound before any balancing is conducted. Optimal refers to the final L1 and biasbound after balance is conducted on the optimal number of dimensions of the left singular vectors of $\K$. The biasbound ratio reports the proportion of the original biasbound to the final optimal biasbound.
\end{table}

\clearpage

\subsection{Performance Across a Range of \emph{b}} \label{app:various_b}
Examining results across a range of values of $b$, we first observe that \textit{kpop} results are highly consistent across a wide range of values for $b$. Notably, at smaller values of $b$, we see how feasibility constraints directly lead to a trade off between achieving calibration on the mean first constraints and seeking balance on the left singular vectors of $\K$. Because these mean first results prioritize mean calibration on all available variables, for small values of $b$, \textit{kpop + mf} struggles to find weights that additionally balance on more than the first handful of left singular vectors of $\K$. This can be seen directly in Table \ref{allB_numdims}. The resulting estimates therefore perform similarly to mean calibration approaches. 

\begin{figure}[!htp]
    \centering
    \includegraphics[width=\textwidth]{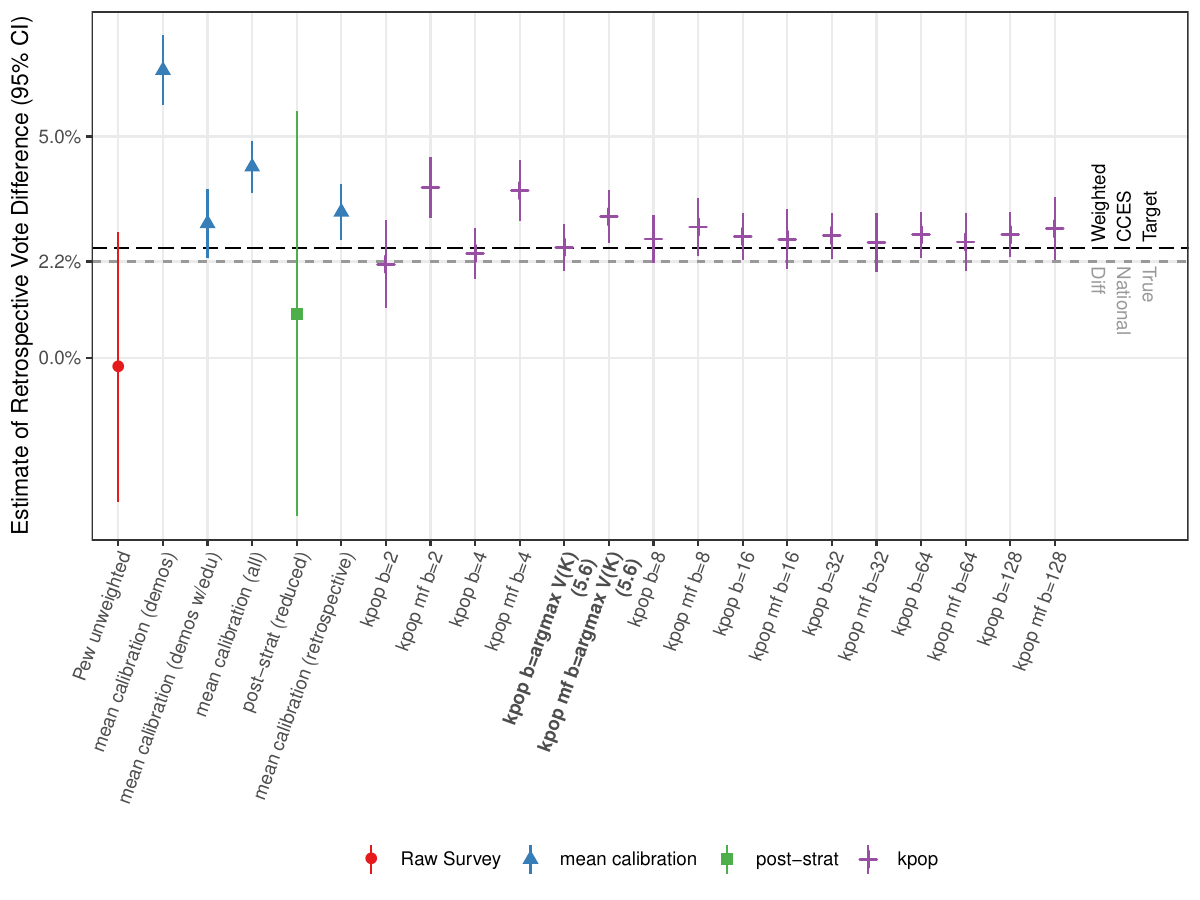}
    \caption{Performance of \kpop\ across a range of choices of $b$ with and without mean first (all) constraints. Note that mean first results append columns of the svd of the raw data and prioritize first achieving mean calibration on these columns before proceeding to balance on the left singular vectors of $\K$.}
\end{figure}

\begin{table}
\caption{Number of Left Singular Vectors of K Balanced on by choice of Gaussian Kernel Bandwidth $b$}\label{allB_numdims}
\begin{center}
\begin{tabular}[t]{rrr}
\toprule
$b$ & \textit{kpop} & \textit{kpop+mf (all)}\\
\midrule
2.00 & 39 & 8\\
4.00 & 159 & 7\\
argmax V(K): 5.66  & 159 & 52\\
8.00 & 168 & 52\\
16.00 & 189 & 101\\
32.00 & 200 & 105\\
64.00 & 203 & 106\\
128.00 & 189 & 50\\
\bottomrule
\end{tabular}
\end{center}
\emph{Note:} \textit{kpop+mf (all)} balances first on 31 dimensions of $svd(X)$ then seeks additional balance on the the number of left singular vectors of $K$ indicated. With small choices of $b$, $\kpop$ struggles to find feasible balance on many dimensions of $svd(K)$ particularly when also seeking balance on $svd(X)$ in the meanfirst approach.
\end{table}

\clearpage

\maketitle

\section{Simulation Details}
\label{app:simulations}

In Section \ref{subsec:sim_intro} we briefly described our simulation
study. While \textit{kpop} achieves good balance and low bias estimates
in our application, we employ a simulation setting to more fully
investigate performance on both bias and variability. To construct a
realistic simulation, we employ the same setting above, but specify the
selection and outcome models to have direct control over the mechanism
of bias.

\hypertarget{probability-sample-selection-model}{%
\subsection{Probability Sample Selection
Model:}\label{probability-sample-selection-model}}

First, we specify a simplistic but non-linear selection model as
follows:

\begin{align*}
p(S=1) &= logit^{-1}\Big( PID(3way) + Age(4way)+ Gender + educ(3way) + Race(4way) \\ 
&+ BornAgain + PID(3way)*Age(4way) + BornAgain*Age(4way)\Big)
\end{align*}

Coefficients are chosen to be roughly similar to a fitted model to pew
that yields a sample size around 500. Namely:

\begin{table}[!h]

\caption{\label{tab:r5_nonlin}Non-Linear Selection Model}
\centering
\begin{tabular}[t]{lr}
\toprule
  & Coefficient Value\\
\midrule
(Intercept) & -2.00\\
pid\_3wayInd & 2.00\\
pid\_3wayRep & 2.00\\
femaleMale & 0.50\\
age\_bucket36 to 50 & 0.15\\
\addlinespace
age\_bucket51 to 64 & 0.20\\
age\_bucket65+ & 0.20\\
educ\_3wayCollege & 0.70\\
educ\_3wayPost-grad & -1.00\\
raceHispanic & 0.50\\
\addlinespace
raceOther & 0.30\\
raceWhite & 0.70\\
bornYes & 2.00\\
age\_bucket36 to 50:recode\_bornYes & 1.00\\
age\_bucket51 to 64:recode\_bornYes & 1.50\\
\addlinespace
age\_bucket65+:recode\_bornYes & 2.00\\
pid\_3wayInd:recode\_age\_bucket36 to 50 & 0.30\\
pid\_3wayRep:recode\_age\_bucket36 to 50 & 0.50\\
pid\_3wayInd:recode\_age\_bucket51 to 64 & 1.00\\
pid\_3wayRep:recode\_age\_bucket51 to 64 & 1.00\\
\addlinespace
pid\_3wayInd:recode\_age\_bucket65+ & -0.20\\
pid\_3wayRep:recode\_age\_bucket65+ & 2.00\\
\bottomrule
\end{tabular}
\end{table}

This yields the following sampling probabilities:

\begin{table}[!h]

\caption{\label{tab:r5_nonlin}Sample Inclusion Probabilities}
\centering
\begin{tabular}[t]{lr}
\toprule
  & Selection Probability\\
\midrule
Min & 0.0008\\
25\% & 0.0065\\
Mean & 0.0134\\
75\% & 0.0158\\
Max & 0.0167\\
\addlinespace
Sum & 513.1844\\
\bottomrule
\end{tabular}
\end{table}

\hypertarget{outcome-model}{%
\subsection{Outcome Model:}\label{outcome-model}}

To keep things straight forward, the outcome model is identical to the
selection model. In other words, again we have:

\begin{align*}
p(Vote=D) &= PID(3way) + Age(4way)+ Gender + educ(3way) + Race(4way) \\
&+ BornAgain + PID(3way)*Age(4way) + BornAgain*Age(4way)
\end{align*}

We add normally distributed noise to this outcome with mean zero and
standard deviation \(\sigma = sd(Y)*1\), yielding an \(R^2\approx.5\).
To yield negative bias, the coefficients in the outcome model start as
the inverse of the coefficients in the selection model, then through an
automated procedure they are adjusted until they produce \(\hat{y}\)'s
that lie within a probability range. This yields a population target in
percentage points of \(\bar{Y} =49.14\%\). The correlation between
selection probability and the probability of voting democratic is
\(\approx -0.6\). This induces a bias in the unweighted sample around
-3.5\%.

\begin{table}[!h]

\caption{\label{tab:r5_nonlin}Non-Linear Selection Mode with R2= 0.5 Outcome Model}
\centering
\begin{tabular}[t]{lr}
\toprule
  & Coefficient Value\\
\midrule
(Intercept) & 0.6250\\
pid\_3wayInd & -0.0750\\
pid\_3wayRep & -0.0750\\
femaleMale & -0.0188\\
age\_bucket36 to 50 & -0.0056\\
\addlinespace
age\_bucket51 to 64 & -0.0075\\
age\_bucket65+ & -0.0075\\
educ\_3wayCollege & -0.0262\\
educ\_3wayPost-grad & 0.0375\\
raceHispanic & -0.0188\\
\addlinespace
raceOther & -0.0112\\
raceWhite & -0.0262\\
bornYes & -0.0750\\
age\_bucket36 to 50:recode\_bornYes & -0.0375\\
age\_bucket51 to 64:recode\_bornYes & -0.0562\\
\addlinespace
age\_bucket65+:recode\_bornYes & -0.0750\\
pid\_3wayInd:recode\_age\_bucket36 to 50 & -0.0112\\
pid\_3wayRep:recode\_age\_bucket36 to 50 & -0.0188\\
pid\_3wayInd:recode\_age\_bucket51 to 64 & -0.0375\\
pid\_3wayRep:recode\_age\_bucket51 to 64 & -0.0375\\
\addlinespace
pid\_3wayInd:recode\_age\_bucket65+ & 0.0075\\
pid\_3wayRep:recode\_age\_bucket65+ & -0.0750\\
\bottomrule
\end{tabular}
\end{table}

\clearpage

\hypertarget{simulation-results}{%
\subsection{Simulation Results:}\label{simulation-results}}

We compare \textit{kpop} and \textit{kpop+mf} against the same set of
mean calibration specifications discussed at length in the application
in section \ref{subsec:appmethods}. This including raking demographics,
demographics with education and raking on all variables. We also compare
performance against post stratification, stratifying on the true sample
selection model. Notably, even though our model is fairly simplistic,
the complexity of the full cross-sectional strata is such that empty
cells post a challenge for post-stratification. On average,
post-stratification must drop around 22.7\% of population units due to
missing strata in the sample. Finally, we include mean calibration on
the true selection model which meats the (link) linear ignorability
assumption as well as the canonical Horvitz-Thompson estimator.

\hypertarget{bias-and-mse}{%
\subsubsection{Bias and MSE}\label{bias-and-mse}}

The resulting bias across a full range of methods is displayed in the
table below.

\begin{table}[!h]

\caption{\label{tab:non_lin_res_race_r55_inter}Simulation Results (arranged by MSE)}
\centering
\begin{tabular}[t]{lrrr}
\toprule
  & Bias (p.p.) & MSE & Abs Bias Reduction\\
\midrule
Unweighted & -3.510 & 12.603 & 0.000\\
mean calibration (demos) & -1.618 & 2.893 & 0.539\\
mean calibration (demos+edu) & -1.296 & 1.961 & 0.631\\
mean calibration (all) & -0.029 & 0.226 & 0.992\\
kpop & -0.272 & 0.357 & 0.923\\
\addlinespace
kpop+mf (demos) & -0.165 & 0.297 & 0.953\\
kpop+mf (demos+edu) & -0.150 & 0.268 & 0.957\\
kpop+mf (all) & 0.012 & 0.244 & 0.997\\
Horvitz-Thompson (true) & -0.160 & 10.229 & 0.954\\
post-stratification (true) & -1.120 & 1.571 & 0.681\\
\addlinespace
mean calibration (true) & -0.010 & 0.214 & 0.997\\
\bottomrule
\end{tabular}
\end{table}

\hypertarget{box-plot-of-estimates-by-method}{%
\subsubsection{Box Plot of Estimates by
Method}\label{box-plot-of-estimates-by-method}}

To see these results visually, we can examine a box plot.

\includegraphics{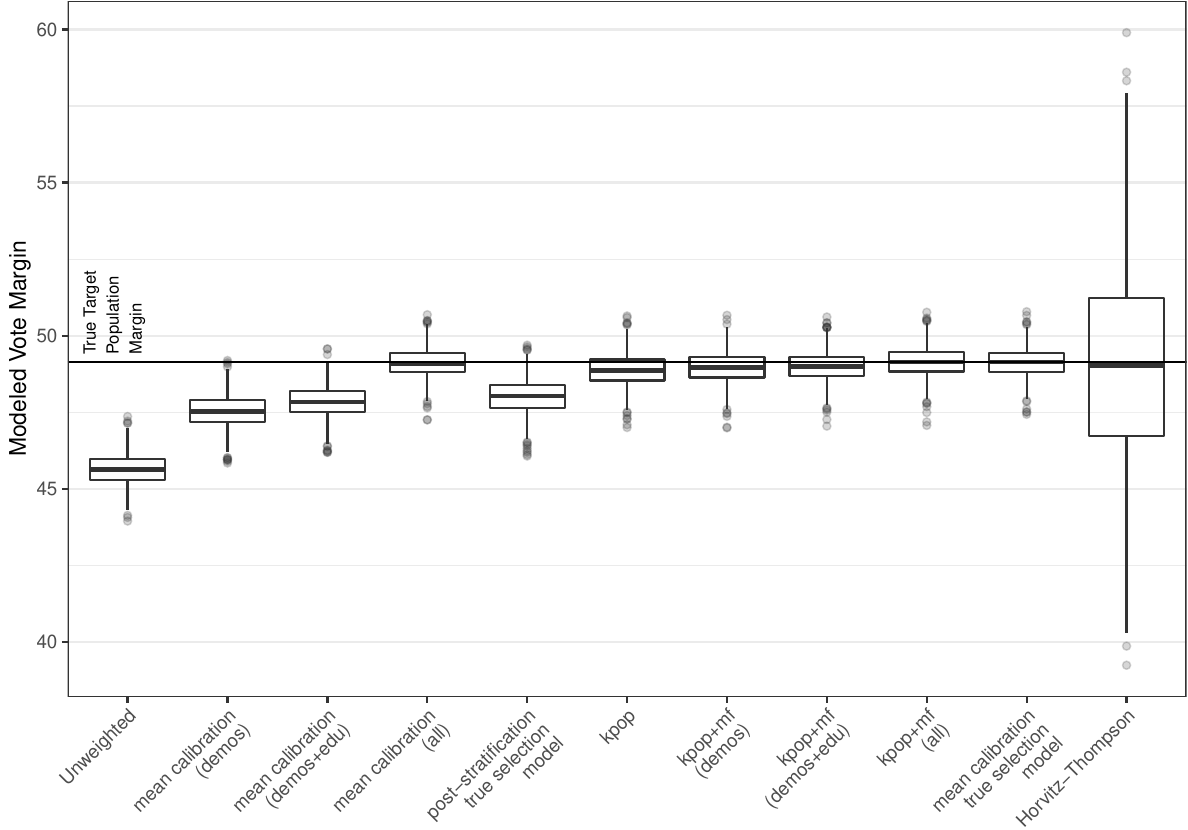}

\hypertarget{standard-errors}{%
\subsubsection{\texorpdfstring{Standard Errors
\label{sims:SEs}}{Standard Errors }}\label{standard-errors}}

The following table present both the empirical standard errors across a
number of estimators that are reviewed and can be easily referenced in
Kott, Phillip S. ``Calibration weighting in survey sampling.'' Wiley
Interdisciplinary Reviews: Computational Statistics 8.1 (2016): 39-53.
These include including SEs that assume fixed weights, SEs that us
linearization variance estimation (eqn 15), and SEs that assume a
quasi-probability sampling process (eqn 14).

\begin{table}[!h]

\caption{\label{tab:SEs}Empirical SE Results \textbf{in Percent} for kpop Methods 1000 sims $R^2$ on Outcome =  0.497}
\centering
\begin{tabular}[t]{lrrrr}
\toprule
  & kpop & \makecell[c]{kpop+mf\\(demos)} & \makecell[r]{kpop+mf\\(d+edu)} & \makecell[l]{kpop+mf\\(all)}\\
\midrule
SE fixed & 0.533 & 0.532 & 0.528 & 0.524\\
SE linear & 0.521 & 0.522 & 0.500 & 0.476\\
SE quasi & 0.522 & 0.520 & 0.499 & 0.474\\
sd(y hat) & 0.533 & 0.520 & 0.496 & 0.494\\
\bottomrule
\end{tabular}
\end{table}

Next, we evaluate the coverage of these various SE estimators and see
all have about nominal or higher coverage.

\begin{table}[!h]

\caption{\label{tab:coverage}Bias-Adjusted SE Coverage Results for kpop Methods 1000 sims $R^2$ on Outcome =  0.497}
\centering
\begin{tabular}[t]{lrrrr}
\toprule
  & kpop & \makecell[c]{kpop+mf\\(demos)} & \makecell[r]{kpop+mf\\(d+edu)} & \makecell[l]{kpop+mf\\(all)}\\
\midrule
SE fixed & 0.951 & 0.955 & 0.957 & 0.964\\
SE linear & 0.943 & 0.954 & 0.948 & 0.946\\
SE quasi & 0.943 & 0.954 & 0.947 & 0.943\\
\bottomrule
\end{tabular}
\end{table}

\hypertarget{weights-diagnostics}{%
\subsubsection{Weights Diagnostics}\label{weights-diagnostics}}

The following table shows the average moments of the weights by method
across 1000 simulations. Note that \textit{"Effective SS"} refers to the
effective sample size calculated using Kish's expression
\(\frac{\left(\sum_i w_i\right)^2}{\sum_i w_i^2}\), but is not well
referenced against a set sample size since we this varied across
simulutions because we used bernoulli draws. The average sample size
should be \(\sum(p(S=1))\) (printed below).
\textit{"No. Units to 90\% Sum of Total"} is the number of units
required to sum to \(90%
\) of the total sum of the weights when weights are ordered from largest
to smallest. In other words, summing from the largest weights to the
smallest, we require this number of units to get to \(90%
\) of the total sum of the weights.

\begin{table}[!h]

\caption{\label{tab:weights}Average Moments of Weights by kpop Method across 1000 simulations}
\centering
\begin{tabular}[t]{lrrrrrr}
\toprule
Estimator & Variance & Max & Min & IQR & \makecell[r]{Effective SS\\(Kish)} & \makecell[l]{No. Units to 90\%\\Sum of Total}\\
\midrule
kpop & 0.590 & 6.033 & 0.167 & 0.688 & 327.670 & 379.044\\
kpop+mf (d+edu) & 0.516 & 5.904 & 0.126 & 0.641 & 341.601 & 387.453\\
kpop+mf (demos) & 0.626 & 6.631 & 0.105 & 0.661 & 319.682 & 379.743\\
kpop+mf (all) & 0.505 & 5.310 & 0.142 & 0.708 & 341.974 & 380.999\\
\bottomrule
\end{tabular}
\end{table}

\hypertarget{dimensions-of-k}{%
\subsubsection{Dimensions of K}\label{dimensions-of-k}}

\begin{table}[!h]

\caption{\label{tab:unnamed-chunk-8}Average Dimensions of K w/ R2= 0.5 Outcome Model}
\centering
\begin{tabular}[t]{lrr}
\toprule
  & Average & SD\\
\midrule
kpop & 47.120 & 23.659\\
kpop+mf (demos) & 28.360 & 12.292\\
kpop+mf (demos+edu) & 16.035 & 7.870\\
kpop+mf (all) & 1.365 & 1.547\\
\bottomrule
\end{tabular}
\end{table}

\end{document}